\documentclass[11pt,a4paper]{article}
\pdfoutput=1
\usepackage{jcappub}

\usepackage{float}
\usepackage{bm}
\usepackage{epsfig}
\usepackage{graphicx}
\usepackage{amsmath}
\usepackage{amssymb}
\usepackage{amsbsy}
\usepackage{color}
\usepackage{subfigure}
\usepackage{slashed}
\usepackage{afterpage}
\usepackage{psfrag}
\usepackage{hyperref}
\usepackage{soul}

\newcommand{\be}{\begin{equation}}
\newcommand{\ee}{\end{equation}}
\newcommand{\bea}{\begin{eqnarray}}
\newcommand{\eea}{\end{eqnarray}}

\renewcommand\({\left(}
\renewcommand\){\right)}
\renewcommand\[{\left[}
\renewcommand\]{\right]}

\newcommand{\GF}{G_{\rm F}}
\newcommand{\CV}{C_{\rm V}}
\newcommand{\CA}{C_{\rm A}}
\newcommand{\bk}{{\bf k}}
\newcommand{\bp}{{\bf p}}
\newcommand{\bq}{{\bf q}}
\newcommand{\opl}{\omega_{\rm p}}

\newcommand{\exclude}[1]{}

\definecolor{gre}{rgb}{0,0.4,0.3}

\begin{document}
\subheader{\hfill MPP-2017-164}

\title{Solar neutrino flux at keV energies}

\author[a]{Edoardo Vitagliano,}
\author[a,b]{Javier~Redondo}
\author[a]{and Georg~Raffelt}

\affiliation[a]{Max-Planck-Institut f\"ur Physik (Werner-Heisenberg-Institut),\\
F\"ohringer Ring 6, 80805 M\"unchen, Germany}

\affiliation[b]{Department of Theoretical Physics, University of Zaragoza, \\ P.\ Cerbuna 12, 50009 Zaragoza, Spain}

\emailAdd{edovita@mpp.mpg.de}
\emailAdd{jredondo@unizar.es}
\emailAdd{raffelt@mpp.mpg.de}

\abstract{We calculate the solar neutrino and antineutrino flux in the
  keV energy range. The dominant thermal source processes are photo
  production ($\gamma e\to e \nu\bar\nu$), bremsstrahlung ($e+Ze\to
  Ze+e+\nu\bar\nu$), plasmon decay ($\gamma\to\nu\bar\nu$), and $\nu\bar\nu$
  emission in free-bound and bound-bound transitions of partially
  ionized elements heavier than hydrogen and helium.
  These latter processes dominate in the energy range of a few keV
    and thus carry information about the solar metallicity.
    To calculate their rate
  we use libraries of monochromatic photon radiative opacities
  in analogy to a previous calculation of solar axion emission. Our
  overall flux spectrum and many details differ significantly from previous
  works. While this low-energy flux is not measurable with present-day
  technology, it could become a significant background for future
  direct searches for keV-mass sterile neutrino dark matter.}

\maketitle

\section{Introduction}
\label{sec:intro}

The nuclear reactions producing energy in the Sun also produce the well-known
solar neutrino flux of about $6.6\times10^{10}~{\rm cm}^{-2}~{\rm s}^{-1}$ with
MeV energies. At Earth this is the
largest neutrino flux, except perhaps in the immediate vicinity of a nuclear
power reactor. The role of solar neutrinos for the discovery of leptonic
flavor conversion and for pioneering the field of astroparticle physics
cannot be overstated. It is a remarkable shift of paradigm that solar
neutrinos today, fifty years after their first detection, are part of the
``neutrino floor,'' the dominant background for direct searches of dark
matter in the form of weakly interacting massive particles (WIMPs).

\begin{figure}[H]
\centering
\includegraphics[width=0.90\textwidth]{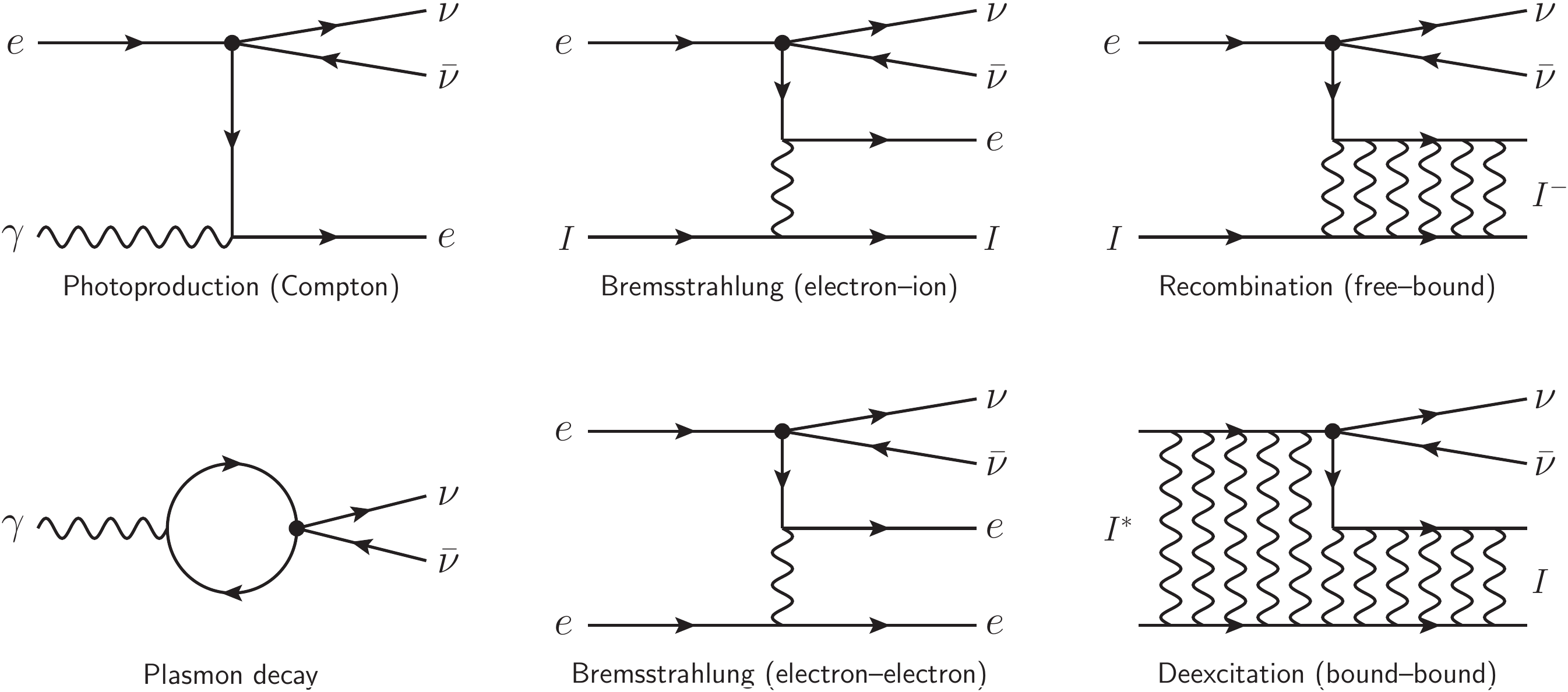}
\caption{Processes for thermal neutrino pair production in the Sun.}
\label{fig:processes}
\end{figure}

Another well-motivated dark matter candidate is a sterile neutrino in the keV
mass range~\cite{Adhikari:2016bei}. One idea for a direct search is the
sterile-neutrino capture on a stable isotope of dysprosium if $m_s>2.83~{\rm
keV}$ \cite{Lasserre:2016eot}. Other searches for slightly heavier sterile neutrinos include unstable isotopes \cite{Li:2010vy,Li:2011mw}, coherent inelastic scattering on atoms \cite{Ando:2010ye} and electron scattering \cite{Campos:2016gjh}. Once again, solar neutrinos could be a
limiting background, now those with keV energies that emerge from various
thermal processes in the solar plasma which has a typical temperature of
1~keV. While this idea is futuristic with present-day technology, it
motivates us to consider keV-range solar neutrinos. This is a standard
neutrino flux, yet it is conspicuously absent from a popular plot of the
``grand unified neutrino spectrum'' at Earth that ranges from cosmic
background neutrinos to those from cosmic-ray sources at EeV energies
\cite{Spiering:2012xe}.\footnote{See also the IceCube MasterClass at
  \url{http://masterclass.icecube.wisc.edu/en/learn/detecting-neutrinos}.} The only detailed previous study of the keV range
solar flux \cite{Haxton:2000xb} ignores bremsstrahlung production and
overestimates photo production by a spurious plasmon resonance. This
situation motivates us to take a completely fresh look, taking advantage
of recent progress in calculating the keV-range solar flux of other low-mass
particles such as axions and hidden photons
\cite{Pospelov:2008jk,Derevianko:2010kz,An:2013yfc,Redondo:2013lna,Redondo:2013wwa,Hardy:2016kme}.

Thermal neutrino emission from stars is an old topic, central to the physics
of stellar evolution, and detailed studies exist as well as Computer routines
to be coupled with stellar evolution codes \cite{Itoh:1996}. However, in this
context neutrinos play the role of a local energy sink for the stellar plasma
and so the emission spectrum is not provided. Moreover, for a low-mass
main-sequence star like our Sun, energy loss by thermal neutrinos is
negligible. Therefore, standard energy-loss rates, which cover a large range
of temperatures, densities and chemical compositions, may not be optimized
for solar conditions.

\begin{figure}[b!]
\centering
\hbox to\textwidth{\includegraphics[height=5.8cm]{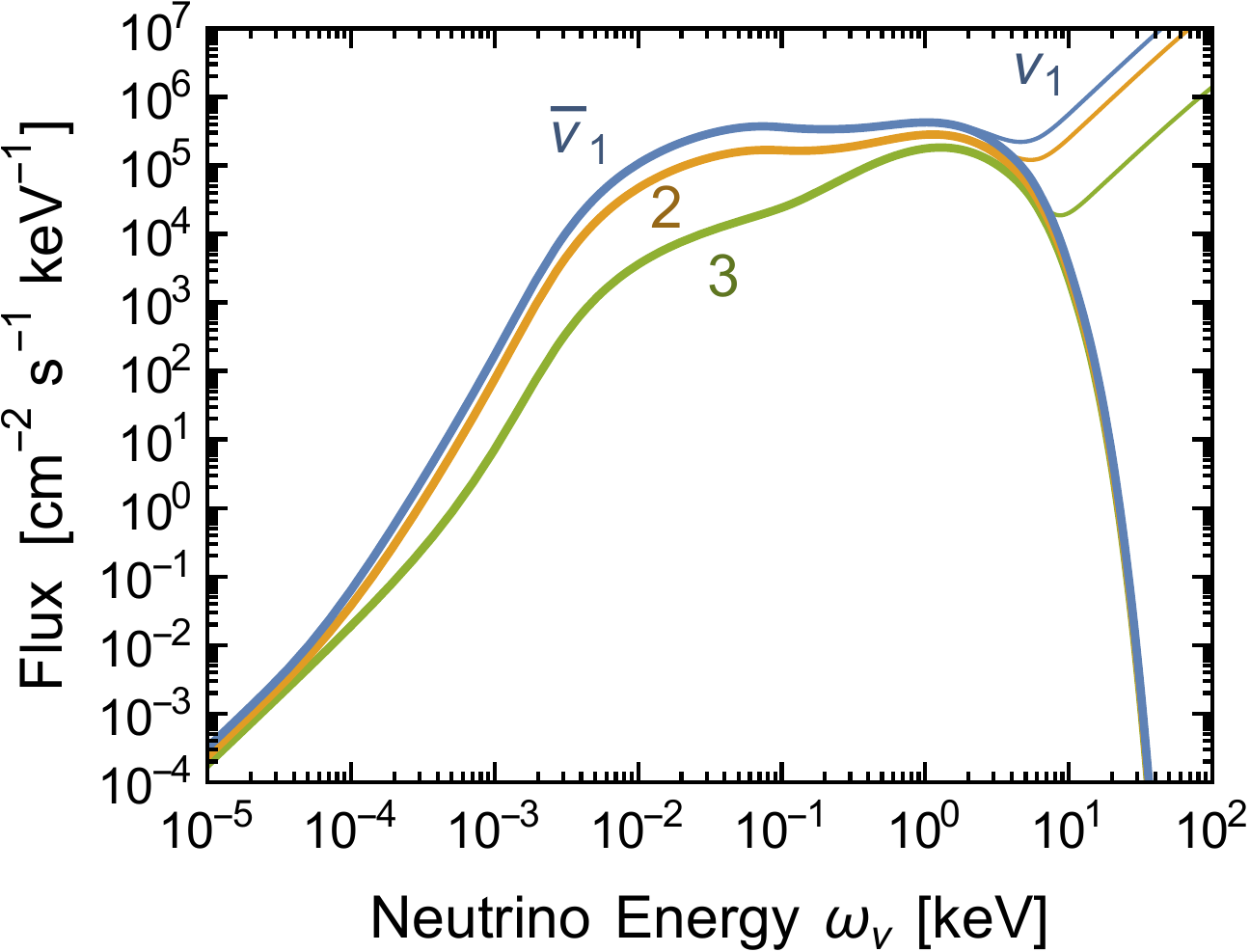}
\hfil\includegraphics[height=5.8cm]{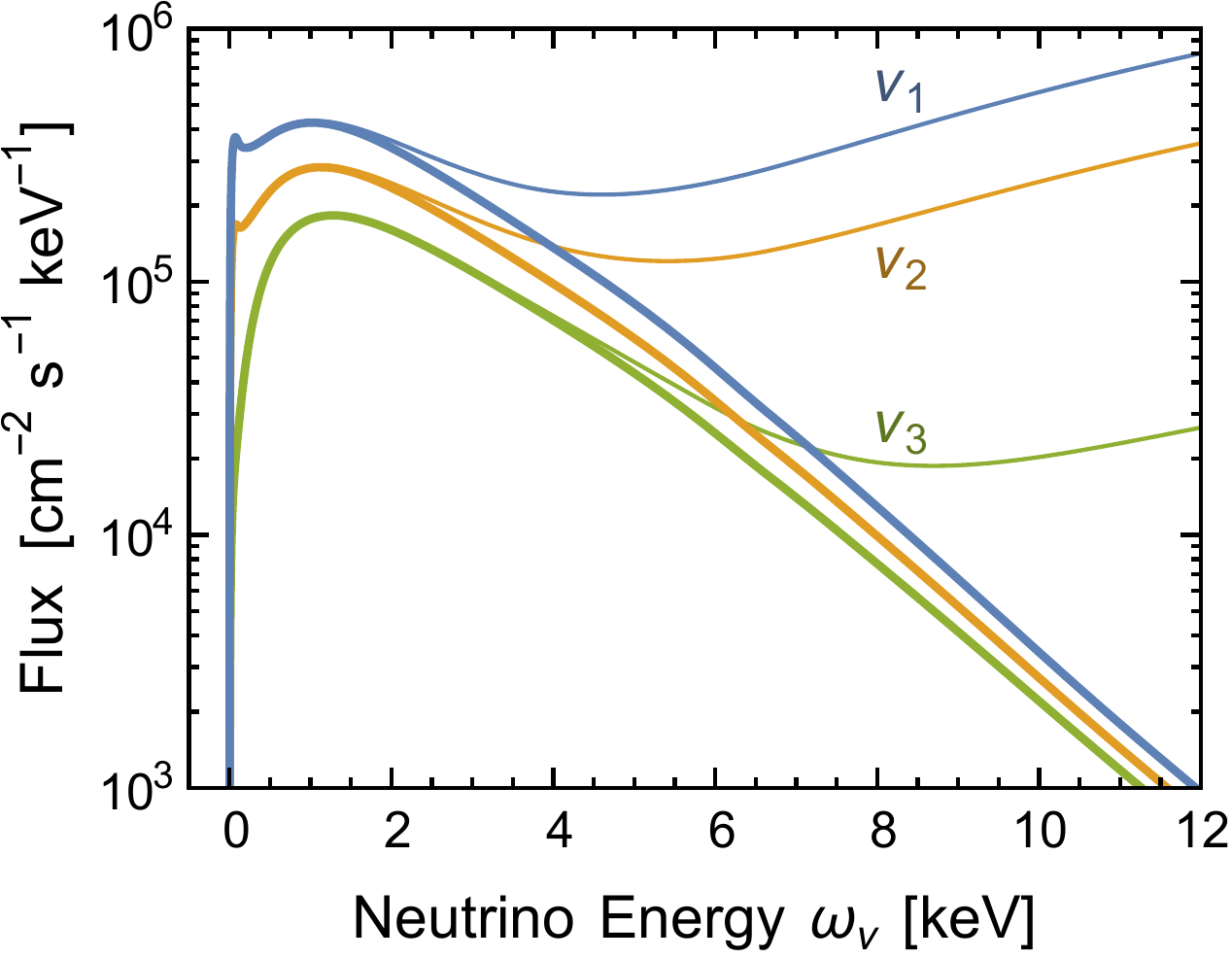}}
\vskip12pt
\hbox to\textwidth{\includegraphics[height=5.8cm]{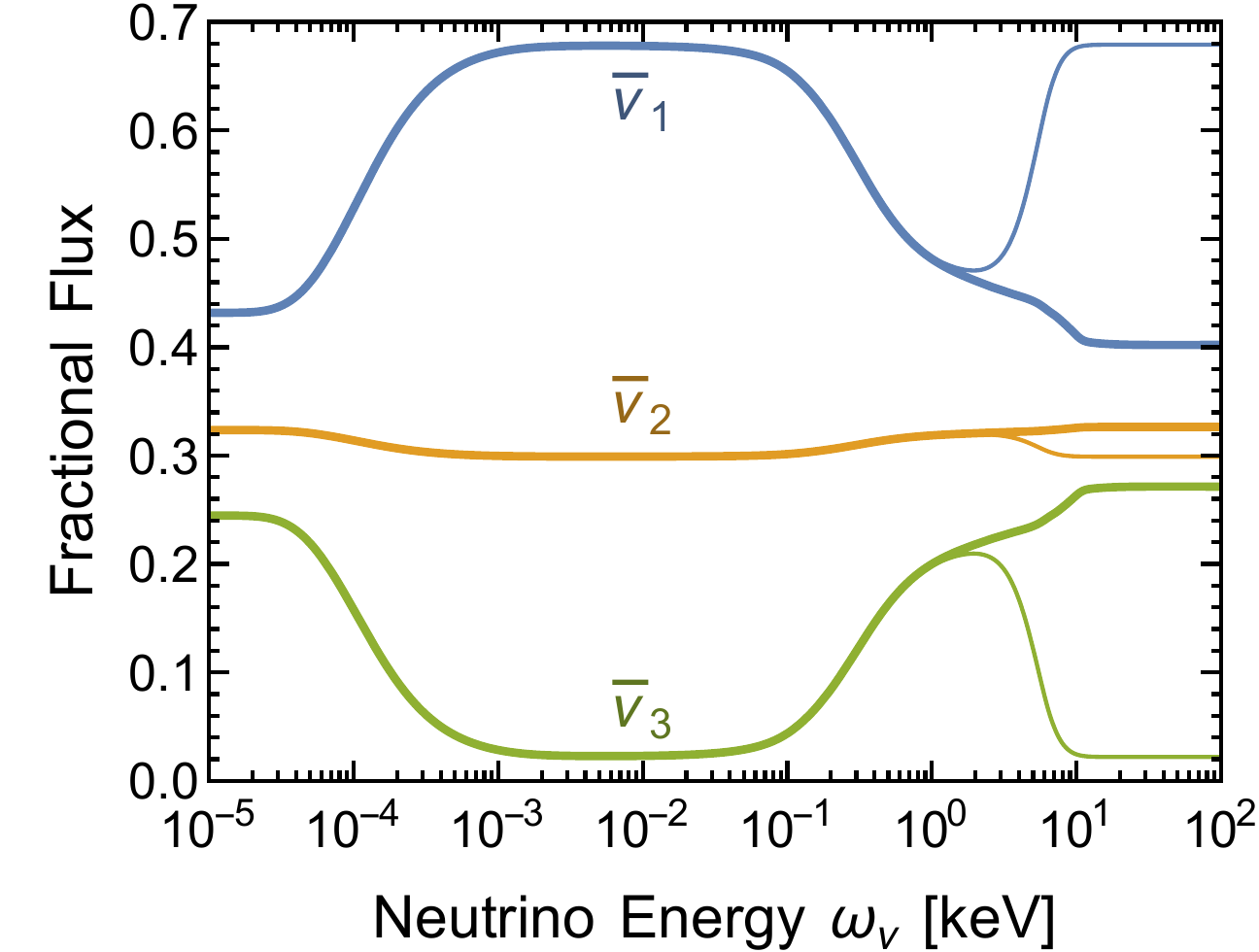}
\hfil\includegraphics[height=5.8cm]{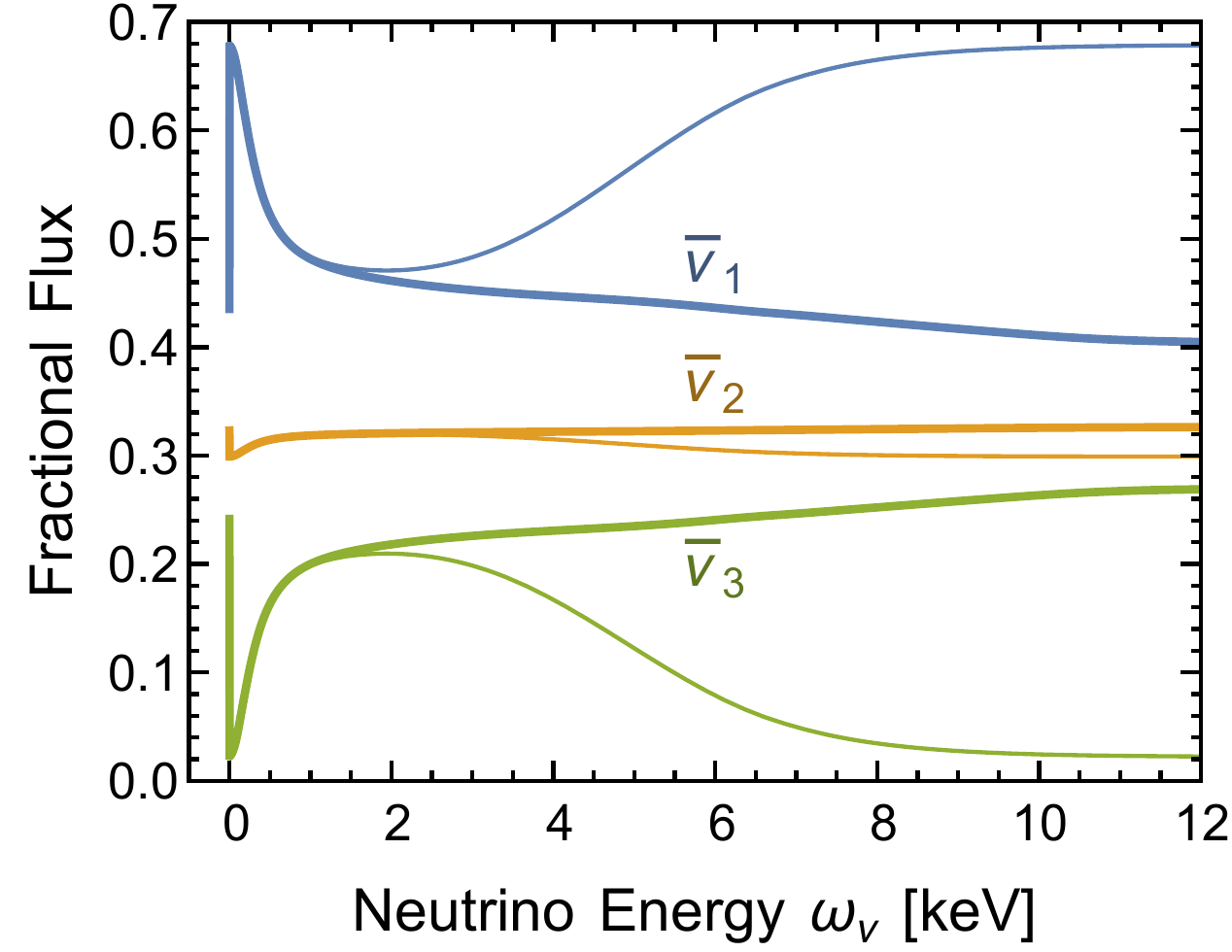}}
\caption{Solar neutrino flux at Earth in the keV range. The flavor
  dependence is given in the mass basis for the 1, 2 and 3 mass
  eigenstates (blue, orange and green). Thick lines are for $\bar\nu$,
  thin lines for $\nu$ which includes a contribution from the nuclear
  pp reaction which produces only $\nu_e$ at the source. The other
  source channels are thermal reactions which produce $\nu$ and
  $\bar\nu$ in equal measure.  The bottom panels show the fractions of
  the total flux provided by the individual mass eigenstates.}
\label{fig:summaryflux}
\end{figure}

Low-energy neutrinos are produced in the solar plasma by the pair-production
processes shown in figure~\ref{fig:processes}, where nonrelativistic
electrons are the sources. Electron velocities and spins are ``kicked'' by the
ambient electromagnetic fields, leading to the emission of neutrino pairs. At
low energies, the weak interaction is sufficiently well described by an
effective four-fermion local interaction
proportional to the Fermi constant $\GF$. The effective coupling constants
for the vector and axial-vector interaction, $\CV$ and $\CA$, are different
for $\nu_e$ and the other flavors, leading to a nontrivial flavor dependence
of the emitted fluxes. The vector-current interaction leads essentially to
electric dipole radiation caused by the time variation of the electron
velocity, whereas the axial-vector current leads to magnetic dipole radiation
caused by fluctuations of the electron spin. Yet in the nonrelativistic
limit, the rates for both mechanisms are related by simple numerical factors
and there is no interference between them, so all processes provide rates
proportional to $(a\, \CV^2+b\,\CA^2)\GF^2$ with coefficients $a$ and $b$
that depend on the specific emission process. One consequence of this simple
structure is that the emission rates are closely related to those for axions
(axial current interaction) or hidden photons (vector current interaction)
and also closely related to photon absorption rates. We will take full
advantage of these similarities, i.e., the relation between these different
processes by simple phase-space factors.

In figure~\ref{fig:summaryflux} we show the overall low-energy solar
neutrino and antineutrino flux at Earth from our calculation.  All
thermal processes shown in figure~\ref{fig:processes} produce $\nu\bar\nu$ pairs
and thus equal fluxes of neutrinos and antineutrinos.
This equipartition is another consequence
of the nonrelativistic approximation, where weak magnetism effects
disappear along with $\CV\CA$ cross terms in the emission
rate~\cite{Horowitz:2001xf}. In addition, the low-energy tail of the
neutrino spectrum produced in the nuclear pp reaction contributes
significantly to the keV flux. At the source, this reaction
produces $\nu_e$ which, like the other channels, have decohered into
their mass components long before they reach Earth. The pp flux causes
an overall asymmetry between the keV-range $\nu$ and $\bar\nu$
spectra. The fractional contribution of the 1, 2 and 3 mass
eigenstates arriving at Earth are shown in the
lower panels of figure~\ref{fig:summaryflux}.
Different emission processes have different energy dependences and there are different coefficients $(a\,\CV^2+b\,\CA^2)$ for $\nu_e$ and the other flavors,
thus explaining the fractional flux variation.

The rest of the paper is devoted to deriving the results shown in
figure~\ref{fig:summaryflux}. In
sections~\ref{sec:plasmon-decay}--\ref{sec:freebound} we study
individual processes. In section~\ref{sec:solarflux} we derive the
overall flux at Earth after integration over a standard solar model
which is detailed in appendix~\ref{app:smm}.
Section~\ref{sec:discussion} is finally devoted to a summary and
discussion.

\section{Plasmon decay}
\label{sec:plasmon-decay}

\subsection{Matrix element}

We begin our calculation of neutrino pair emission from the solar
interior with plasmon decay, $\gamma\to\nu\bar\nu$, the process of
figure~\ref{fig:processes} involving the smallest number of
participating particles. This process is also special in that it has
no counterpart for axion emission.
In any medium, electromagnetic excitations with
wave vector $k=(\omega,\bk)$ acquire a nontrivial dispersion relation
that can be written in the form $\omega^2-\bk^2=\Pi(\bk)$, where
$\Pi_\bk=\Pi(\bk)$ is the on-shell polarization function. We will always
consider an unmagnetized and isotropic plasma. It supports both
transverse (T) modes, corresponding to the usual photons, and
longitudinal (L) modes, corresponding to collective oscillations of
electrons against ions. Whenever $\Pi_\bk>0$ (time like dispersion),
the decay into a neutrino pair, taken to be massless, is kinematically
allowed. For both T and L modes, neutrino pairs are actually emitted
by electrons which oscillate coherently as a manifestation of the
plasma wave.

\begin{figure}[htbp]
\begin{center}
\includegraphics[width=6cm]{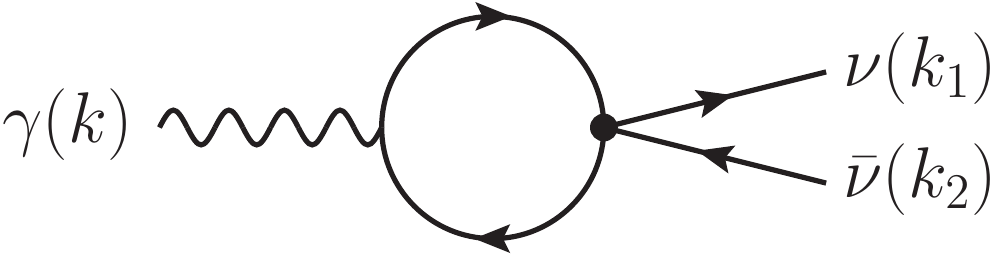}
\caption{Plasmon decay mediated by electrons of the medium.}
\label{fig:plasmon-graph}
\end{center}
\end{figure}

Therefore, plasmon decay and actually all other processes relevant for thermal
pair emission in the Sun depend on the neutrino-electron interaction.
At low energies, it is given by the effective
neutral-current interaction
\begin{equation}\label{eq:NC-interaction}
  {\mathcal L}_{\rm int}=\frac{\GF}{\sqrt{2}}\,
  \bar\psi_e\gamma^\mu(\CV-\CA\gamma_5 )\psi_e\,
  \bar\psi_\nu\gamma_\mu(1-\gamma_5)\psi_\nu\,,
\end{equation}
where $\GF$ is Fermi's constant. The effective vector (V) and axial-vector (A)
coupling constants include a neutral-current contribution and for
$\nu_e$ also a charged-current piece from $W^\pm$ exchange. Altogether
one finds
\begin{subequations}
\begin{eqnarray}
  \hbox to 7.cm{$\CV=\frac{1}{2}(4\sin\Theta_{\rm W}+1)$
    \quad and\quad$\CA=+\frac{1}{2}$\hfil}&&
  \hbox to 3.cm{for\quad$\nu_e$,\hfil}\\
  \hbox to 7.cm{$\CV=\frac{1}{2}(4\sin\Theta_{\rm W}-1)$
    \quad and\quad$\CA=-\frac{1}{2}$\hfil}&&
  \hbox to 3.cm{for\quad$\nu_\mu$ and $\nu_\tau$,\hfil}
\end{eqnarray}
\end{subequations}
where $4\sin^2\Theta_{\rm W}=0.92488$ in terms of the weak mixing
angle.
In particular, this implies that the rates of A processes,
proportional to $\CA^2=1/4$, are the same for all flavors. On the
other hand, the rates for V processes are proportional to
\begin{equation}
   \CV^2=0.9263~~\hbox{for}~\nu_e\bar\nu_e
   \quad\hbox{and}\quad
   \CV^2=0.0014~~\hbox{for}~\nu_{\mu,\tau}\bar\nu_{\mu,\tau}\,.
\end{equation}
Thus for heavy-lepton neutrinos we may
safely ignore the vector-current interaction, i.e., such processes
produce an almost pure $\nu_e\bar\nu_e$ flux.

Plasmon decay has been extensively
studied in the literature \cite{Adams:1963zzb,Zaidi:1963zzb,Braaten:1993jw,Haft:1993jt,Ratkovic:2003td}.
The squared matrix element for the
transition $\gamma\to\nu\bar\nu$ with photon four-momentum
$k=(\omega,\bk)$ and $\nu$ and $\bar\nu$ four momenta
$k_1=(\omega_1,\bk_1)$ and $k_2=(\omega_2,\bk_2)$ is found to be
(see figure~\ref{fig:plasmon-graph}),
\begin{equation}
  |\mathcal{M}_{\gamma\to\nu\bar\nu}|^2=
  \frac{\CV^2\GF^2}{8\pi\alpha}\,Z_{\bk}\Pi^2_{\bk}\,
  \epsilon_\mu\epsilon_\nu^* N^{\mu\nu},
\end{equation}
where $\alpha=e^2/4\pi$ is the fine-structure constant.
$Z_\bk$ is the on-shell
wave-function renormalization factor and
$\Pi_\bk$ the polarization factor appropriate for the T or L
excitation. The photon polarization vector is $\epsilon^\mu$
with $\epsilon^\mu\epsilon_\mu^*=-1$. The neutrino tensor, appearing in all
pair emission processes, is
\begin{equation}\label{eq:neutrino-tensor}
  N^{\mu\nu}=8\bigl(k_1^\mu k_2^\nu+k_1^\nu k_2^\mu
  -k_1{\cdot}k_2\,g^{\mu\nu}+i\varepsilon^{\alpha\mu\beta\nu}k_{1\alpha}k_{2\beta}\bigr).
\end{equation}
Inserting this expression in the squared matrix element yields
\begin{equation}
  |\mathcal{M}_{\gamma\to\nu\bar\nu}|^2=
  \frac{\CV^2\GF^2}{\pi\alpha}\,Z_{\bk}\Pi^2_{\bk}\,
  (\epsilon^*{\cdot}k_1\,\epsilon{\cdot}k_2+\epsilon{\cdot}k_1\,\epsilon^*{\cdot}k_2+k_1{\cdot}k_2\bigr)\,.
\end{equation}
Notice that the axial-vector interaction does not induce plasmon decay under the approximations described. This is
particularly obvious in the nonrelativistic limit where we can think
of the emission process as dipole radiation from coherently
oscillating electrons, whereas the electron spins, responsible for
non-relativistic axial-current processes, do not oscillate coherently. The
absence of a sizeable axial-current rate implies that
plasmon decay produces with high accuracy only $\nu_e\bar\nu_e$ pairs.

\subsection{Nonrelativistic limit}

In a classical plasma (nonrelativistic and nondegenerate), the
electromagnetic dispersion relations for transverse (T) and
longitudinal (L) plasmons are found to be
\begin{equation}
\omega^2\big|_{\rm T}=
    \opl^2\left(1+\frac{\bk^2}{\opl^2+\bk^2}\,\frac{T}{m_e}\right)+\bk^2
\qquad\hbox{and}\qquad
  \omega^2\big|_{\rm L}=\opl^2\left(1+3\,\frac{\bk^2}{\opl^2}\,\frac{T}{m_e}\right).
\end{equation}
The plasma frequency is given in terms of the electron density
$n_e$ by
\begin{equation}
    \opl^2=\frac{4\pi\alpha\,n_e}{m_e}\,.
\end{equation}
In the Sun, $T\lesssim 1.3~{\rm keV}$ so that $T/m_e\lesssim 0.0025$ and
with excellent approximation we may limit our discussion to the
lowest-order term. Moreover, the lowest-order expression pertains
to any level of degeneracy as long as the electrons remain nonrelativistic.
In this case, T modes propagate in the same way
as particles with mass $\opl$, i.e., $\omega^2=\bk^2+\opl^2$, whereas
L modes oscillate with a fixed frequency $\omega=\opl$, independently
of $\bk$.
Therefore, the L-plasmon dispersion relation is
time-like only for $|\bk|<\opl$, so only these soft quanta can decay
into neutrino pairs.

In the nonrelativistic limit and using Lorentz gauge one finds
$Z_{\rm T}=1$, $\Pi_{\rm T}=\opl^2$,
$Z_{\rm L}=\opl^2/(\opl^2-\bk^2)$ and
$\Pi_{\rm L}=\opl^2-\bk^2$. Without loss of generality,
we may assume the photon to move in the $z$ direction. The T
polarization vectors are in this case $\epsilon^\mu=(0,1,0,0)$ and
$\epsilon^\mu=(0,0,1,0)$, respectively, whereas the L case with
$|\bk|<\opl$ has $\epsilon^\mu=(|\bk|,0,0,\opl)/(\opl^2-\bk^2)^{1/2}$.
In Coulomb gauge one finds different expressions for the L quantities.

\subsection{Decay rate and spectrum}

Next we consider the decay rate of a transverse or longitudinal
on-shell plasmon with wave vector $\bk$ and ask for its decay rate
\begin{equation}\label{eq:Gamma-plasmon}
  \Gamma_{\gamma\to\nu\bar\nu}=
  \int\frac{d^3\bk_1}{(2\pi)^3}\,\frac{d^3\bk_2}{(2\pi)^3}\,
  \frac{|{\cal M}_{\gamma\to\nu\bar\nu}|^2}{2\omega\,2\omega_1\,2\omega_2}\,
  (2\pi)^4\,\delta^4(k-k_1-k_2)\,.
\end{equation}
In a nonrelativistic plasma one easily finds the usual result
\begin{equation}
  \Gamma_{\rm T}=\Gamma_{\rm p}\,\frac{\opl}{\omega_\bk}
    \qquad\hbox{and}\qquad
  \Gamma_{\rm L}=\Gamma_{\rm p}\,\frac{(\opl^2-\bk^2)^2}{\opl^4}
   \qquad\hbox{with}\qquad
 \Gamma_{\rm p}=\frac{\CV^2\GF^2\opl^5}{48\,\pi^2\alpha}\,.
\end{equation}
For T plasmons $\omega_\bk=(\opl^2+\bk^2)^{1/2}$
with $0\leq|\bk|<\infty$, whereas for L
plasmons the decay is allowed for $0\leq|\bk|<\opl$. The T case
is reminiscent of a decaying particle with mass $\opl$ where the laboratory decay rate
is time-dilated by the factor $\opl/\omega_\bk$. In the limit $\bk\to 0$ both rates are the same.
Indeed, in the limit of vanishing wave number one cannot distinguish a transverse from
a longitudinal excitation.

We are primarily interested in the neutrino energy spectrum. The symmetry
of the squared matrix element under the exchange $k_1\leftrightarrow k_2$ implies
that it is enough to find the $\nu$ spectrum which is identical to the one for $\bar\nu$.
Therefore, in equation~(\ref{eq:Gamma-plasmon}) we integrate over $d^3\bk_2$ to remove
the momentum delta function, and over $d\Omega_1$ to remove the one for energy conservation.
Overall, with $\omega_\nu=\omega_1$ we write the result in the form
\begin{equation}
\frac{d\Gamma}{d\omega_\nu}=\Gamma\,g(\omega_\nu)\,,
\end{equation}
where $g(\omega_\nu)$ is a normalized function. For T
plasmons, averaged over the two polarization states, we find
\begin{equation}\label{eq:T-spectrum}
g_{\rm T}(\omega_\nu)= \frac{3}{4}\,\frac{\bk^2+\(\omega_\bk-2\,\omega_\nu\)^2}{|\bk|^3}
\quad\hbox{for}\quad \frac{\omega_\bk-|\bk|}{2} <\omega_\nu< \frac{\omega_\bk+|\bk|}{2}
\end{equation}
and zero otherwise. If the T plasmon were an unpolarized massive
spin-1 particle, this would be a top-hat spectrum on the shown
interval, corresponding to isotropic emission boosted to the
laboratory frame. However, the T plasmon misses the third polarization
state so that even unpolarized T plasmons do not show this
behavior. For L plasmons we find
\begin{equation}
g_{\rm L}(\omega_\nu)= \frac{3}{2}\,\frac{\bk^2-\(\opl-2\,\omega_\nu\)^2}{|\bk|^3}
\quad\hbox{for}\quad \frac{\opl-|\bk|}{2} <\omega_\nu< \frac{\opl+|\bk|}{2}
\end{equation}
and zero otherwise, with the additional constraint $0\leq |\bk|<\opl$.

\begin{figure}[htbp]
\begin{center}
\includegraphics[width=7.5cm]{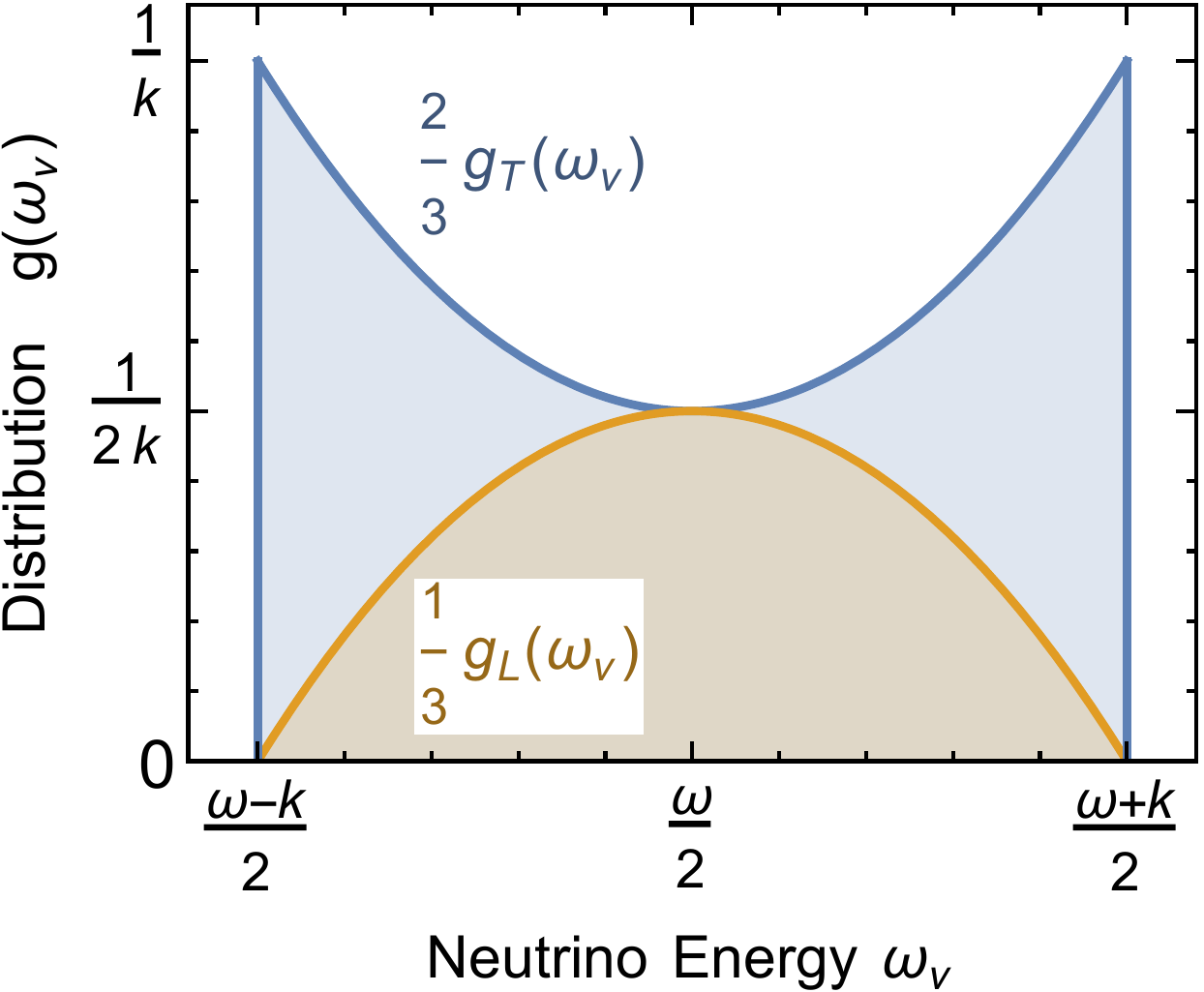}
\caption{Normalized $\nu$ spectrum from transverse and longitudinal
plasmon decay $\gamma\to\nu\bar\nu$. For T plasmons $\omega=(\opl^2+\bk^2)^{1/2}$,
whereas for L plasmons $\omega=\opl$.}
\label{fig:plasmon-spectra}
\end{center}
\end{figure}

We show these distributions in figure~\ref{fig:plasmon-spectra}.
Assuming equal $\omega$ for both types of excitations and also
equal $\bk$, the distributions add to a top-hat spectrum of the form
$\frac{2}{3}\,g_{\rm T}(\omega_\nu)+\frac{1}{3}\,g_{\rm L}(\omega_\nu)=1/|\bk|$
on the interval
$(\omega-|\bk|)/2<\omega_\nu<(\omega+|\bk|)/2$, i.e.,
this average resembles the decay spectrum of an unpolarized spin-1
particle.
However, the dispersion relations are different for T and L plasmons
so that, for equal $\bk$, they have different $\omega$ and these
distributions are not on the same $\omega_\nu$ interval. The only
exception is the $\bk\to0$ limit when $\omega\to\opl$ for both
types.

\subsection{Thermal emission spectrum}

\begin{figure}[b!]
\centering
\hbox to \textwidth{\includegraphics[height=5.7cm]{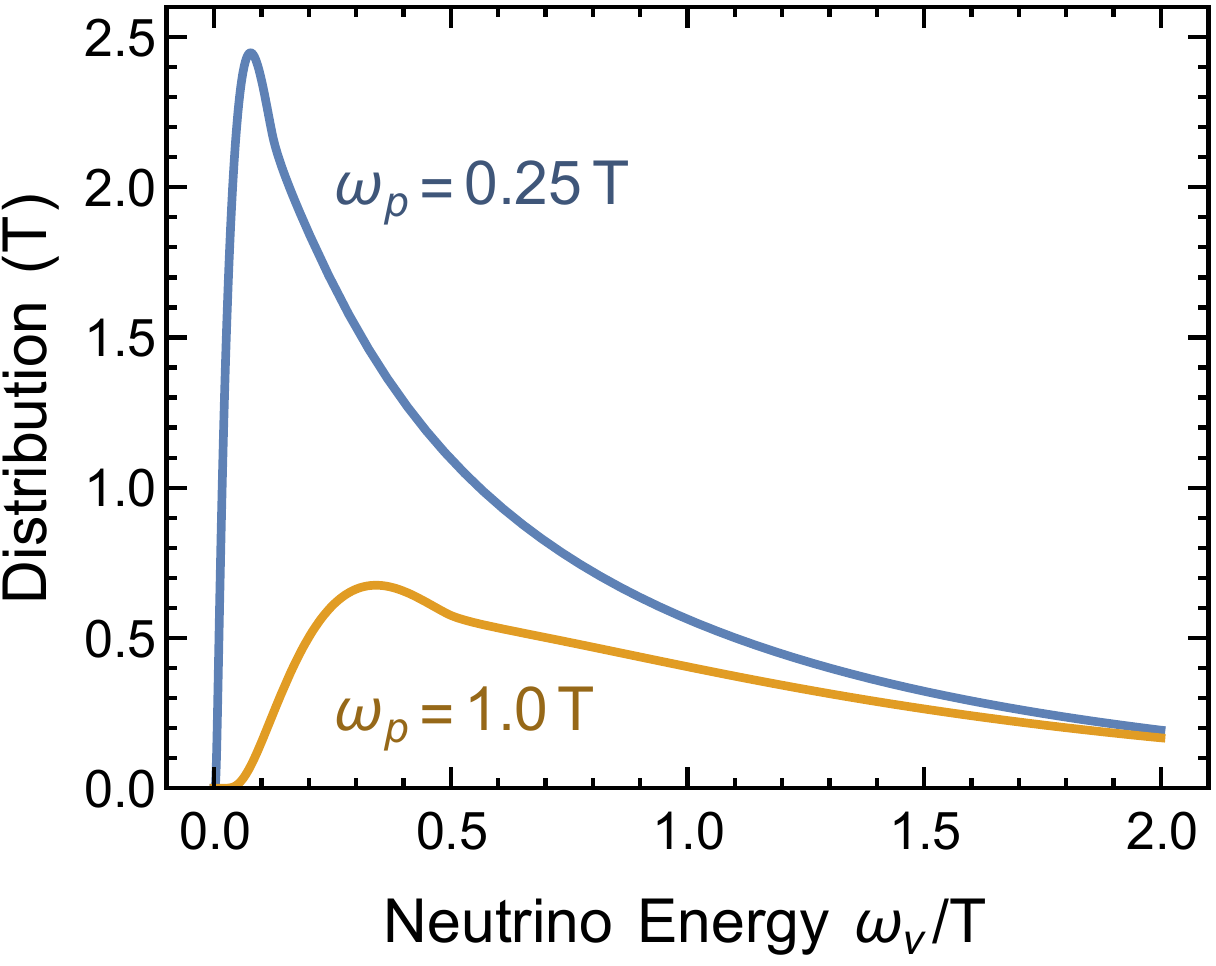}\hfil\includegraphics[height=5.7cm]{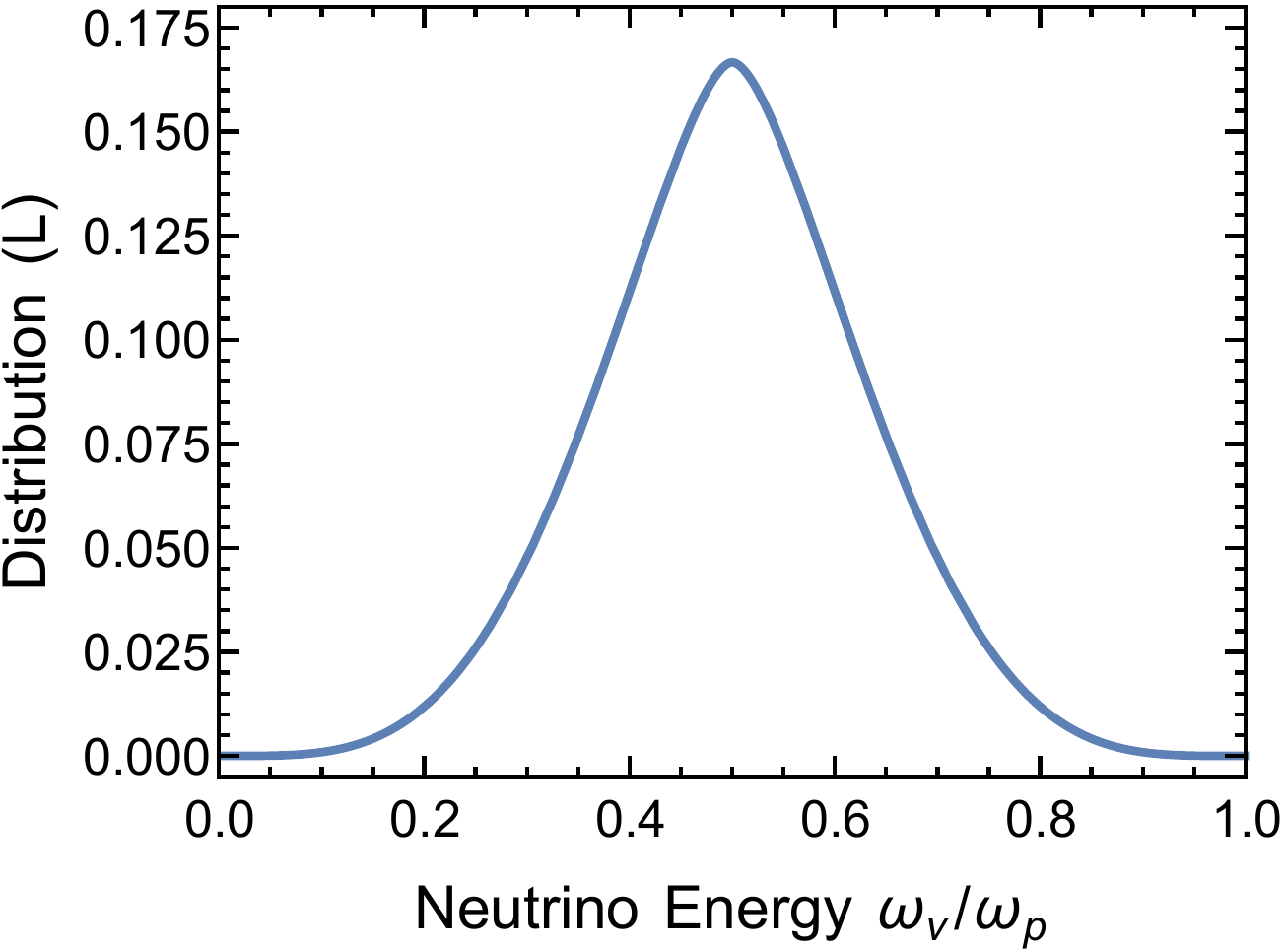}}
\caption{Neutrino spectrum from thermal plasmon decay.
  {\em Left panel:\/} Transverse plasmons. The curves represent the dimensionless integral in equation~(\ref{eq:T-spectrum-final}) and correspond to $\opl/T=0.25$ and 1 as indicated.
  {\em Right panel:\/} Longitudinal plasmons. The curve is the dimensionless integral in equation~(\ref{eq:L-spectrum}). To make the vertical scale comparable to T plasmons, a factor
  $(\opl/T)/(e^{\opl/T}-1)=1+\mathcal{O}(\opl/T)$ must be included.}
\label{fig:TL-spectrum}
\end{figure}

As our final result we determine the spectral emission density from a
nonrelativistic plasma with temperature $T$. The number of neutrinos emitted
per unit volume per unit time per unit energy interval from T plasmon decay is
\begin{equation}\label{eq:T-spectrum-final}
\frac{d \dot n_\nu}{d\omega_\nu}\Big|_{\rm T}=
\int\limits_{V_\bk}\frac{d^3\bk}{(2\pi)^3}\,\frac{2 \Gamma_{\rm T} g_{\rm T}(\omega_\nu)}{e^{\omega_\bk/T}-1}
=\frac{3\,\Gamma_{\rm p}\opl T}{4\pi^2}\!\!\int\limits_{\omega_\nu+\frac{\opl^2}{4\omega_\nu}}^{~\infty}
\!\!\frac{d\omega}{T}\,\frac{1}{e^{\omega/T}-1}\,\[1+\frac{(\omega-2\omega_\nu)^2}{\omega^2-\opl^2}\]\,.
\end{equation}
The integration is over the volume in $\bk$-space allowed by the decay kinematics
given in equation~(\ref{eq:T-spectrum}) and the factor of 2
accounts for two transverse degrees of freedom.
For $\omega_\nu\ll\opl$ the required plasmon energy
is large so that we may approximate $e^{\omega/T}-1\to e^{\omega/T}$ and
$(\omega-2\omega_\nu)^2/(\omega^2-\opl^2)\to 1$. In this case the
dimensionless integral is \smash{$2 e^{-\opl^2/4\omega_\nu T}$},
i.e., this neutrino flux is exponentially suppressed at low energies due to the exponential suppression of
the density of T-plasmons with sufficient energy.
In figure~\ref{fig:TL-spectrum} we show
the dimensionless integral as a function of $\omega_\nu/T$
for $\opl/T=0.25$ and 1. Notice that in the central solar region
$T=1.3~{\rm keV}$ and $\opl=0.3~{\rm keV}$ so that
$\opl/T=0.25$ corresponds approximately to conditions of the central
Sun. The external shells of the Sun, where $\opl^2/T$ is smaller, turn out to be
relevant for the lowest energy neutrinos from T-plasmon decay. However, we show later that this contribution is subdominant.

For L plasmons, the integral over the initial photon distribution
yields the spectrum of the number emission rate
\begin{eqnarray}\label{eq:L-spectrum}
\frac{d \dot n_\nu}{d\omega_\nu}\Big|_{\rm L}&=&
\int\limits_{V_\bk}\frac{d^3\bk}{(2\pi)^3}\,\frac{\Gamma_{\rm L}
  g_{\rm L}(\omega_\nu)}{e^{\opl/T}-1}
\nonumber\\
&=&\frac{3\Gamma_{\rm p}\,\opl^2}{4\pi^2(e^{\opl/T}-1)}\,
\int\limits_{|\opl-2\omega_\nu|}^{~\opl}
\!\!d|\bk|\,\frac{|\bk|\,\(\opl^2-\bk^2\)^2}{\opl^6}
\,\[1-\frac{(\opl-2\omega_\nu)^2}{\bk^2}\]
\nonumber\\[1ex]
&=&\frac{3\Gamma_{\rm p}\,\opl T}{4\pi^2}\,\frac{\opl/T}{e^{\opl/T}-1}\,\,
\(\frac{2+3y^2-6y^4+y^6}{12}+y^2\log|y|\)\,,
\end{eqnarray}
where $y=(2\omega_\nu-\opl)/\opl$ equivalent to
$\omega_\nu=(y+1)\,\opl/2$. L plasmons have the fixed
energy $\opl$, yet neutrinos from decay occupy the full
interval $0<\omega_\nu<\opl$ owing to the peculiar L dispersion
relation. The neutrino distribution is symmetric relative to
$\omega_\nu=\opl/2$. In figure~\ref{fig:TL-spectrum}
we show the dimensionless $\omega_\nu$ distribution which
is universal for any value of $\opl$.

\begin{figure}[htbp]
\centering
\includegraphics[width=7.5cm]{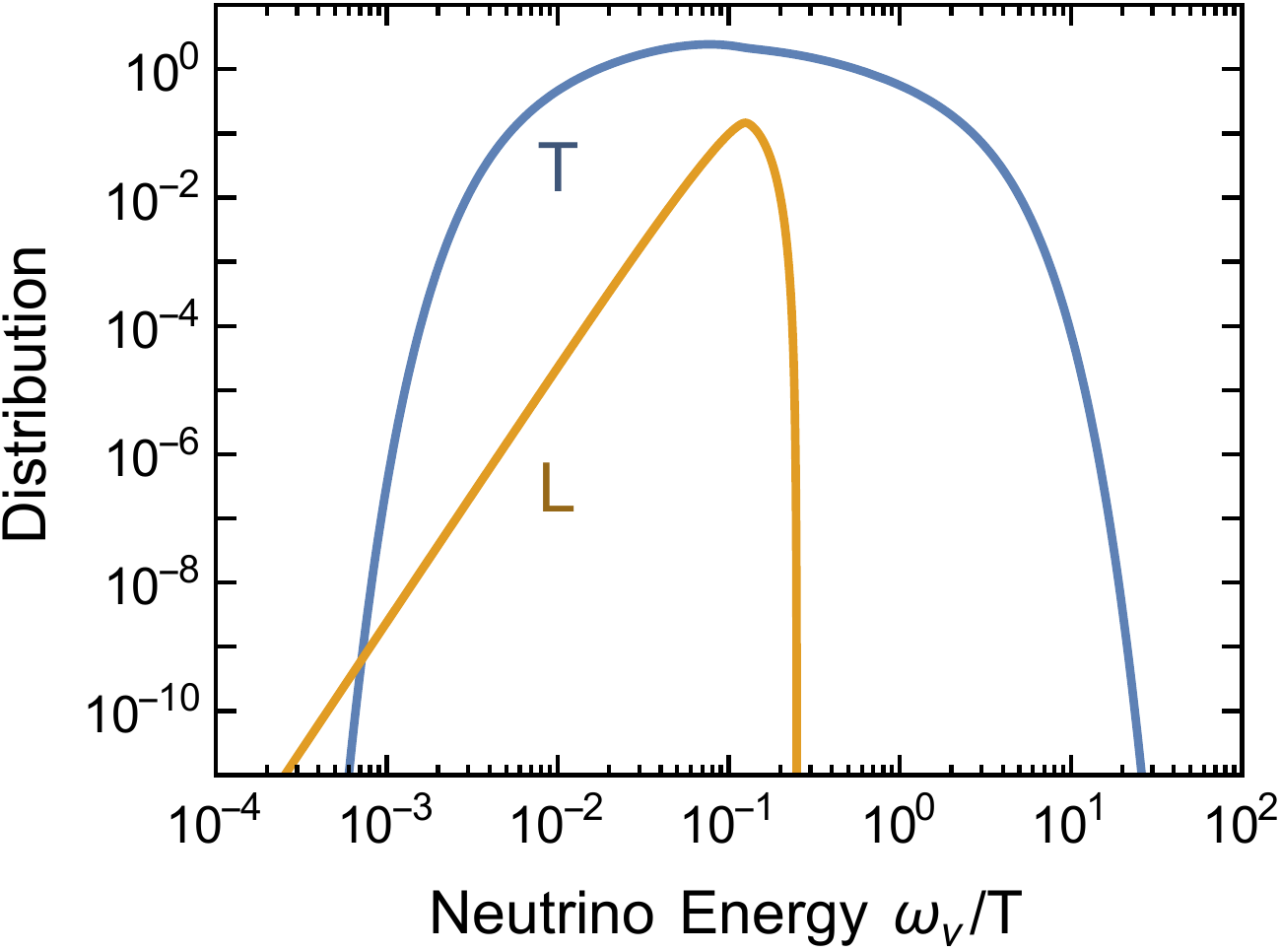}
\caption{Neutrino spectrum from thermal T and L plasmon decay for
$\opl/T=0.25$ in units of $3\Gamma_{\rm p}\opl T/4\pi^2$. At
very low energies, L-plasmon decay dominates.}
\label{fig:plasmon-log-spectrum}
\end{figure}

For $\omega_\nu\ll\opl$ the dimensionless integral can be expanded
and yields $(32/3)\,(\omega_\nu/\opl)^4$, i.e., the spectrum decreases as a power law.
Because the T-plasmon spectrum decreases exponentially,
the L-plasmon decay provides the dominant
neutrino flux at very low $\omega_\nu$.
We illustrate this point in figure~\ref{fig:plasmon-log-spectrum}
where we show both spectra in common units of $3\Gamma_{\rm p}\opl T/4\pi^2$
for $\opl/T=0.25$ on a log-log-plot. The central solar temperature is 1.3~keV,
so L-plasmon decay takes over for sub-eV neutrinos where the overall rate is
extremely small.

\subsection{Compton pole process?}
\label{sec:pole}

Thus far we have used kinetic theory in that we treat the excitations
of the medium as free particles which propagate until they decay or
collide. Plasmons were treated as quasi-stable excitations,
distributed as an ideal Bose gas, which occasionally decay into a
neutrino pair. In a previous study of solar thermal neutrino emission
\cite{Haxton:2000xb} another channel was considered in the form
$\gamma+e^-\to e^-+\gamma$ followed by $\gamma\to\nu\bar\nu$, i.e.,
the decaying plasmon was treated as an intermediate virtual particle
in a Compton-like process. In its propagator, an imaginary part (a
width) was included, but it was stressed that this width is small and
that one needs to integrate over a narrow range of virtual
energy-momenta near the on-shell condition. This ``Compton plasmon
pole'' process was found to dominate thermal pair emission, a finding
that would change everything about neutrino energy losses from stars.

However, we think this result is spurious. In a plasma, of course any
particle is an intermediate state between collisions and as such a
pole in a more complicated process. It is the very basis of the
kinetic approach to treat particles as on-shell states coming from far
away without memory of their previous history. This assumption need
not always be justified, but there is no particular reason why in the
present context it should not apply to the plasmon. Its width is very
small as stressed in reference~\cite{Haxton:2000xb}.

On the other hand, it is not wrong to trace the plasmon one step back
in its collision history. In this case one must be consistent,
however, as to which processes produce and absorb plasmons and are
thus responsible for its width. In reference~\cite{Haxton:2000xb} the
plasmon width was taken as a complicated expression from the
literature based essentially on inverse bremsstrahlung, whereas the
last thing the plasmon did before decaying was taken to be Compton
scattering. In this way, their equation~(11) includes in the numerator
essentially the Compton production rate, in the denominator the
imaginary part of the propagator based on inverse bremsstrahlung. The
emission rate gets spuriously enhanced by a large ratio of plasmon
interaction rates based on different processes.

In summary, as long as the plasmon width is small, as everybody agrees
it is, the ``pole process'' is identical with the plasmon decay
process, not a new contribution. The only difference is that for a
given momentum, the plasmon energy distribution is taken to follow a
delta function (plasmon decay) or a narrow resonance distribution
(pole process). The overall normalization is the same in both cases.

\subsection{Solar neutrino flux}

\begin{figure}[htbp]
\centering
\hbox to\textwidth{\includegraphics[height=5.8cm]{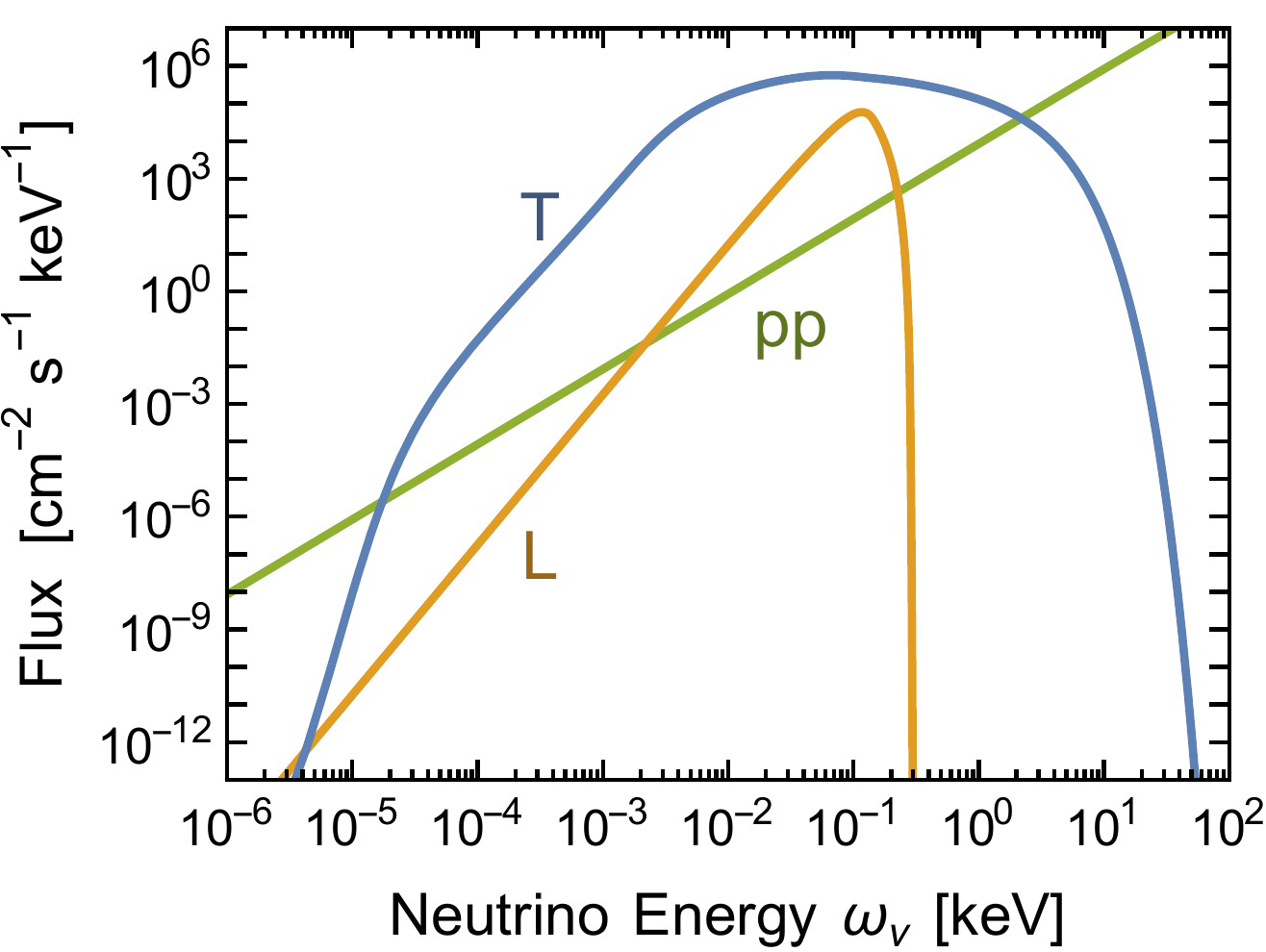}\hfil
\includegraphics[height=5.8cm]{pics/fig07a.pdf}}
\caption{Solar neutrino flux at Earth from transverse and longitudinal plasmon decay. This is the total $\nu$ flux produced as nearly pure $\nu_e$ in the Sun. There is an equal $\bar\nu$ flux. We also show the
  low-energy tail of the usual flux from the nuclear pp reaction which are
  born as $\nu_e$.}
\label{fig:plasma-flux}
\end{figure}

We finally integrate the plasmon decay rate over a standard solar
model and show the expected neutrino flux at Earth in
figure~\ref{fig:plasma-flux}. We specifically use a solar model
from the Saclay group which is described in more detail in
appendix~\ref{app:smm}. At this stage we do not worry about
flavor oscillations and simply give the all-flavor flux at Earth,
recalling that plasmon decay produces pure $\nu_e\bar\nu_e$ pairs
at the source. We find
\begin{equation}
\Phi_{\rm T}=4.12\times10^5~{\rm cm}^{-2}~{\rm s}^{-1}
\qquad\hbox{and}\qquad
\Phi_{\rm L}=4.67\times10^3~{\rm cm}^{-2}~{\rm s}^{-1}
\end{equation}
for the integrated fluxes at Earth.

The T plasmon flux now reaches much smaller energies than
in figure~\ref{fig:plasmon-log-spectrum} when we considered
conditions near the solar center. Very low-energy
neutrinos from plasmon decay require the plasmon to be very
relativistic because the accessible energy range is
$\omega-|\bk|<\omega_\nu<\omega+|\bk|$, so the low energy flux is
exponentially suppressed due to the exponential suppression of the density of
high-energy plasmons. At larger solar radii $T$ is
smaller, but $\opl$ drops even faster
and T plasmons are more relativistic. Therefore,
lower-energy neutrinos become kinematically allowed, i.e.,
lower-energy neutrinos derive from larger solar radii. From
figure~\ref{fig:plasma-flux} we conclude that the L plasmon flux
begins to dominate at energies so low that the assumption of
massless neutrinos is not necessarily justified.

For comparison we also show the $\nu$ flux from the pp reaction
that produces the lowest-energy flux from nuclear reactions.
All standard solar models agree on this flux within around 1\%,
so we may use as a generic number $6.0\times10^{10}~{\rm cm}^{-2}~{\rm s}^{-1}$.
The reaction $p+p\to d+e^++\nu_e$ has
a neutrino endpoint energy of $Q=420.22~{\rm keV}$. Including the thermal
kinetic energy of the protons in the solar plasma, the overall endpoint of
the solar spectrum is $Q=423.41~{\rm keV}$ \cite{Bahcall:1997eg}. This reference
also gives a numerical tabulation of the solar $\nu_e$ spectrum from the pp reaction.
Ignoring a small
$e^+$ final-state correction, the spectrum follows that of an allowed
weak transition of the normalized form
\begin{eqnarray}
\frac{dN}{d\omega_\nu}&=&
\frac{\omega_\nu^2\,(Q+m_e-\omega_\nu)\sqrt{(Q+m_e-\omega_\nu)^2-m_e^2}}{A^5}
\nonumber\\[1ex]
&=&\frac{(Q+m_e)\sqrt{Q(Q+2m_e)}}{A^5}\,\omega_\nu^2+{\cal O}(\omega_\nu^3)
\end{eqnarray}
where $A=350.8~{\rm keV}$ if we use the solar endpoint energy.
The analytic form of the normalization factor $A$ is too complicated to be shown here.
The low-energy pp flux spectrum at Earth is the power law
\begin{equation}\label{eq:ppflux}
\frac{d\Phi_{pp}}{d\omega_\nu}=8150~{\rm cm}^{-2}~{\rm s}^{-1}~{\rm keV}^{-1}\,\(\frac{\omega_\nu}{\rm keV}\)^2\,,
\end{equation}
an approximation that is good to about $\pm1.5$\% for energies below 10~keV.
This shallow power law is simply determined by the low-energy neutrino phase space.
It dominates over plasmon decay at very low energies, although, of course, it does
not produce antineutrinos. We will show later that both are subdominant compared to the neutrino flux produced in bremsstrahlung transitions.

\section{Photo production}
\label{sec:compton}

\subsection{Matrix element and decay rate}

\begin{figure}[htbp]
\centering
\includegraphics[width=6cm]{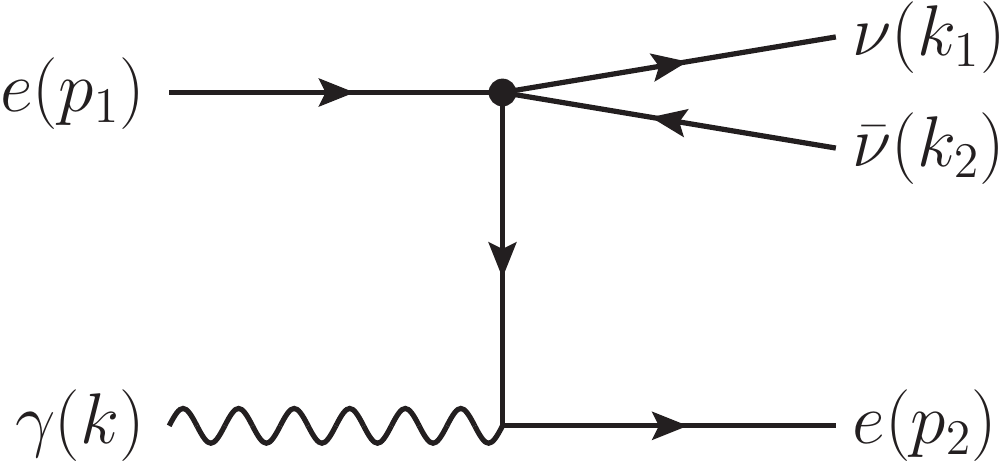}
\caption{Photo production of neutrino pairs (Compton process).
  A second diagram with vertices interchanged is not shown.
}
\label{fig:compton-graph}
\end{figure}

The Compton process (figure~\ref{fig:compton-graph}), also known as
photo production or photoneutrino production, was one of the first
processes to be considered as an energy-loss mechanism for stars
\cite{Chiu:1961zza, Petrosian:1967alk, Ritus:1961alk}. In these older
papers, only the energy-loss rate was calculated, whereas the neutrino
spectrum was calculated in the nonrelativistic limit in
reference~\cite{Haxton:2000xb} and for general kinematics in
reference~\cite{Dutta:2003ny}. We restrict ourselves to the
nonrelativistic limit where electron recoils are neglected. The
process then amounts to the conversion $\gamma\to\nu\bar\nu$,
catalyzed through bystander electrons which take up three-momentum,
and as such is somewhat similar to plasmon decay.  However, no plasma
mass is needed because momentum is taken up by the electrons. Even
though we neglect recoils, the process is not ``forward'' for the
electrons. We can interpret plasmon decay as the coherent version of
photo production.

In the nonrelativistic limit we find for the squared matrix element, averaged over initial spins
and polarizations and summed over final ones,
\begin{equation}
\overline{|\mathcal{M}|^2}=\frac{1}{4}\sum_{\epsilon,s_1,s_2}|\mathcal{M}|^2
=\frac{e^2 \GF^2}{\omega^2}\, M^{\mu\nu}N_{\mu\nu}
\end{equation}
where the neutrino tensor was given in equation~(\ref{eq:neutrino-tensor}). The nonrelativistic
electron tensor for the Compton process is~\cite{Haxton:2000xb}
\begin{equation}
M^{\mu\nu}=\sum_\epsilon\Bigl\lbrace\(\CV^2+\CA^2\)\(-\omega\epsilon^\mu+\delta^{\mu0}\epsilon{\cdot}q\)
\(-\omega\epsilon^\nu+\delta^{\nu0}\epsilon{\cdot}q\)
+\CA^2\[k^\mu k^\nu-(\epsilon{\cdot}q)^2 g^{\mu\nu}\]\Bigr\rbrace\,,
\end{equation}
where $\epsilon$ is the photon polarization vector and $q=k_1+k_2$ the four momentum carried away by
the neutrino pair.

It is noteworthy that both the vector and axial-vector currents contribute
on comparable levels, in contrast to plasmon decay. This is heuristically understood if we observe that the vector part amounts
to electric dipole emission by the electron being ``shaken'' by the incoming EM wave. The rate is
proportional to $(\GF/m_e)^2$ because the outgoing radiation couples with strength $\GF$ and the mass appears due to its inertia against the acceleration. Axial-current
emission amounts to magnetic dipole emission caused by the electron spin. The coupling is through the electron dipole moment $\propto 1/m_e$, so the rate is also
proportional to $(\GF/m_e)^2$. In the case of axion emission,
$\gamma+e\to e+a$, enabled by a derivative axial-vector coupling, the
rate is suppressed by a factor $(\omega/m_e)^2$ relative to Compton
scattering. For neutrinos the coupling structure is the same for both
axial and vector coupling. Note however that these considerations require some handwaving and one should always check which terms in the nonrelativistic expansion of the Hamiltonian contribute to a certain order.

We use the symmetry under the exchange $1\leftrightarrow 2$ and
integrate over the phase space of $\bar\nu$ and over the angles of
$\nu$. With $\opl=0$ we find for the differential ``decay rate'' of T
plasmons with energy $\omega$ of either polarization
\begin{equation}\label{comptonrate}
\frac{d\Gamma_\omega}{d\omega_\nu}= n_e\, \frac{2}{3}\,\frac{\GF^2 \alpha}{\pi^2 m_e^2}\, \(\CV^2+5\CA^2\) \,\frac{(\omega-\omega_{\nu})^2 \omega_{\nu}^2}{\omega}\[1-\frac{2}{3}\frac{(\omega-\omega_{\nu})\omega_{\nu}}{\omega^2}\]
\quad\hbox{for}\quad \omega_\nu< \omega \,.
\end{equation}
Integrated over the photon distribution it gives the familiar result~\cite{Haxton:2000xb}
\begin{eqnarray}\label{eq:comptonrate-2}
  \frac{d\dot n_{\nu}}{d\omega_\nu}&=& n_e\frac{2}{3}\,\frac{\GF^2 \alpha}{\pi^4 m_e^2} \,
  \int_{\omega_\nu}^\infty d\omega\,
  \omega_\nu^2(\omega-\omega_\nu)^2\,\frac{1}{e^{\omega/T}-1}\nonumber\\[2ex]
&&\kern5em{}\times
  \(\CV^2+5\CA^2\) \,\omega\[1-\frac{2}{3}\frac{(\omega-\omega_{\nu})\omega_{\nu}}{\omega^2}\]
\,.
\end{eqnarray}
Including the modified dispersion relation in the plasma with a
  nonvanishing $\opl$ leads to a more complicated expression that modifies the result for $\omega$ near $\opl$ and by up to
  a few percent elsewhere, for us a negligible correction.  On the
  other hand, at energies near $\opl$, Compton emission is subdominant
  relative to plasmon decay. Moreover, one should then also worry
about longitudinal plasmons which can be understood as collective
electron oscillations. One therefore would need to avoid double
counting between $\gamma_{\rm L}+e\to e+\nu\bar\nu$ and bremsstrahlung
$e+e\to e+e+\nu\bar\nu$, see the related discussion in
reference~\cite{Raffelt:1987np}. Therefore, we use the emission
  rate based on the $\opl=0$ expression of
  equation~\eqref{eq:comptonrate-2}, but we will include $\opl$
  in the phase space of initial T plasmons, cutting off $\omega<\opl$
  intial-state photons.

\subsection{Correlation effects}
\label{sec:correlatoins}

So far we have assumed that electrons are completely uncorrelated and
the overall neutrino emission rate is the incoherent sum from
individual scattering events. However, electrons are anticorrelated
by the Pauli exclusion principle and by Coulomb repulsion, both
effects meaning that at the location of a given electron it is less
likely than average to find another one. These anticorrelations lead
to a reduction of the rate, i.e., we need to include a structure
factor $S(\bq^2)$ where $\bq=\bk-\bk_1-\bk_2$ is the
three-momentum transfer. For photon transport, exchange effects
produce a 7\% correction in the solar center and less elsewhere,
whereas Coulomb correlations provide a 20--30\% correction~\cite{Boercker:1987jn}.

Beginning with the exchange correlation, a simple approach is to
include a Pauli blocking factor $(1-f_\bp)$ for the final state
electron in the phase-space integration. For nonrelativistic electron
targets that barely recoil, the final-state $\bp$ can be taken the
same as the initial one, so the overall reduction is the average Pauli
blocking factor
\begin{equation}\label{eq:Reta-definition}
  R_\eta=\frac{2}{n_e}\,\int\frac{d^3\bp}{(2\pi)^3}\,f_\bp(1-f_\bp)
=\int_0^\infty \frac{dx\,x^2}{e^{x^2-\eta}+1}\(1-\frac{1}{e^{x^2-\eta}+1}\)
  \Bigg/\int_0^\infty \frac{dx\,x^2}{e^{x^2-\eta}+1}\,,
\end{equation}
where the nonrelativistic degeneracy parameter $\eta=(\mu-m_e)/T$ is given by
\begin{equation}\label{eq:electron-density}
n_e=2\int\frac{d^3\bp}{(2\pi)^3}\,\frac{1}{e^{\frac{\bp^2}{2m_eT}-\eta}+1}\,.
\end{equation}
We have checked that this expression is indeed the $|\bq|=\kappa\to 0$ limit of
$1+h_x(\kappa)$ given in equation~(6) of reference
\cite{Boercker:1987jn}. In this paper and the literature on solar
opacities, the Pauli blocking effect is interpreted as an
anticorrelation of the electrons in analogy to what is caused
by a repulsive force. So we
can picture Pauli effects either as a blocking of final electron states in collisions
or as an anticorrelation of initial-state electron targets.

Next we turn to Coulomb repulsion where for solar conditions the
structure function can be reasonably approximated essentially
by a Debye-H\"uckel screening prescription \cite{Boercker:1987jn}
\begin{equation}\label{eq:S-Coulomb}
S_e(\bq^2)=1-\frac{k_e^2}{\bq^2+k_e^2+k_i^2}\,.
\end{equation}
The screening scales are
\begin{equation}\label{eq:screening-scales}
k_e^2=R_\eta\,\frac{4\pi\alpha n_e}{T}
\qquad\hbox{and}\qquad k_i^2=\frac{4\pi\alpha}{T}\sum_Z Z^2 n_Z\,,
\end{equation}
where $n_Z$ is the number density of ions with charge $Z e$. For
electrons, we have included the correction factor $R_\eta$ for partial
degeneracy.\footnote{Our $R_\eta$ defined in
  equation~(\ref{eq:Reta-definition}) is the same as $R_\alpha$ defined
  in reference~\cite{Boercker:1987jn} by a ratio of Fermi integrals. The
  overall structure factor was written in the form
  $1+h_x(\kappa)+h_r(\kappa)$ with $h_r(\kappa)=-R_\eta
  k_e^2/(\kappa^2+k_e^2+k_i^2)$. However, for the exchange
  correlations, the $\kappa\to0$ limit is justified and
  $R_\eta=1+h_x(0)$ so that $1+h_x+h_r=R_\eta-R_\eta
  k_e^2/(\kappa^2+k_e^2+k_i^2)=R_\eta S_e(\kappa)$. In other words,
  the exchange correlations indeed amount to a global factor $R_\eta$ for
  final-state Pauli blocking besides a reduction of the electron
  screening scale $k_e^2$.}
For conditions of the central Sun we have an electron density of about
$n_e=6.3\times10^{25}~{\rm cm}^{-3}$ and a temperature $T=1.3~{\rm keV}$,
providing $\eta=-1.425$, leading to $R_\eta=0.927$. The Debye-H\"uckel
scales are $k_e=5.4~{\rm keV}$ and $k_i=7.0~{\rm keV}$.
As these scales are comparable to a typical momentum
transfer there is no simple limit for Coulomb corrections.
In our numerical estimate of the solar emission we use a simple prescription
to account for this effect: the largest possible momentum transfer for
an initial photon of energy $\omega$ is $|\bq_{\rm max}|=2\omega$. Using
$S_e(4\omega^2)$ to multiply  equation~(\ref{comptonrate})
provides an upper limit to the suppression caused by
Coulomb correlations, i.e., the neutrino flux will be slightly overestimated.

Coulomb correlations apply to processes where the electron density
is the crucial quantity, i.e., to the vector-current
part proportional to $\CV^2$. The axial-current contributions, proportional to
$\CA^2$, depend on the electron spins which are not correlated by
Coulomb interactions. If an electron at a given location
has a certain spin, the chance of finding one with the same or opposite spin
at some distance is the same, i.e., the spins are not correlated. Therefore,
the interference of spin-dependent scattering amplitudes from different
electrons average to zero and we do not need any Coulomb correlation correction.
Only the emission of $\nu_e\bar\nu_e$ has any V contribution and
in all cases, the A term strongly dominates. Therefore, overall
Coulomb corrections are small for photo production.

Treating exchange corrections as an average final-state Pauli blocking
factor reveals that it applies for both V and A processes. We can also see
this point in terms of initial-state correlations. Electrons of opposite spin
are not correlated because they can occupy the same location, whereas those
with equal spin ``repel'' each other. Therefore, the interference
effects between initial electrons of equal and opposite spins do not
average to zero.

In summary, the photo production rate is reduced by the overall Pauli
blocking factor $R_\eta$ given in equation~(\ref{eq:Reta-definition})
which in the Sun is a few percent.
The terms propoportional to $\CV^2$, on the other hand, require the
additional Coulomb structure factor given in
equation~(\ref{eq:S-Coulomb}) which can be a 30\% correction.
The V channel essentially applies only to $\nu_e\bar\nu_e$ emission,
where Coulomb correlations provide an overall reduction of perhaps
10\%.

\subsection{Solar neutrino flux}

\begin{figure}[htbp]
\centering
\includegraphics[height=5.8cm]{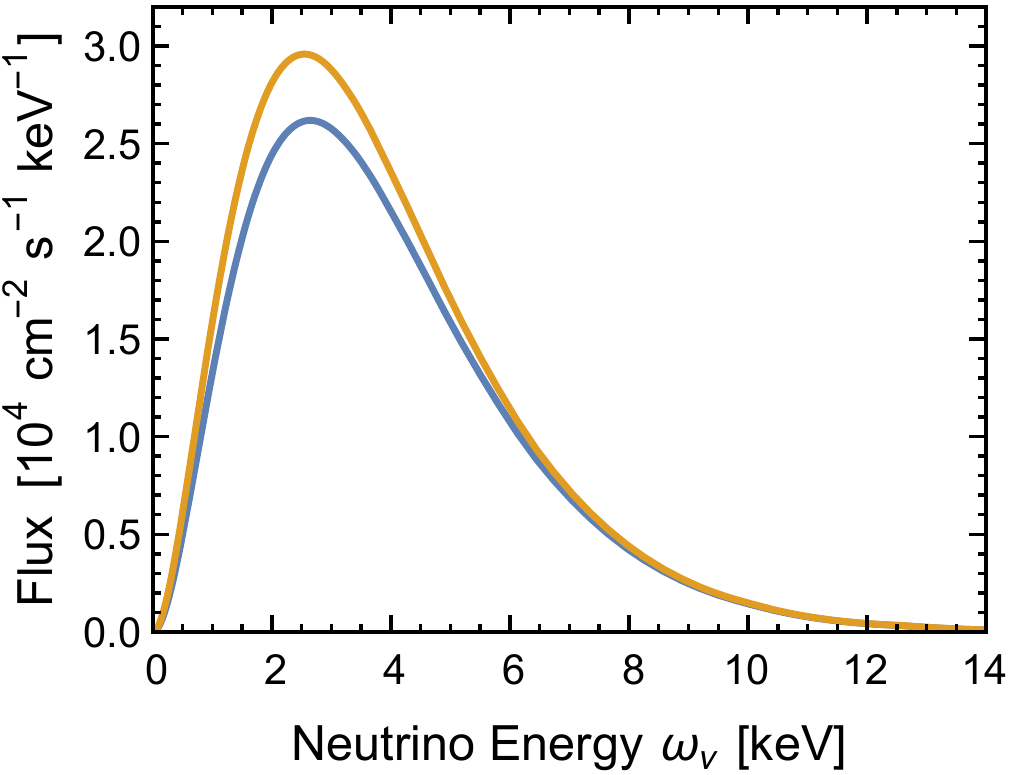}
\caption{Neutrino flux from Compton production for the vector (blue) and
axial-vector (orange) interaction. For the proper flux, the V curve is to be
multiplied with $\CV^2$, the A curve with $5\CA^2$. The difference between
the blue and orange curves derives from Coulomb correlations which apply
only to the V channel. The Coulomb correlations were treated in an approximate
way as described in the text and the suppression could be slightly larger.
}
\label{fig:comptonflux1}
\end{figure}

We now integrate the source reaction rate over our solar model and show the
neutrino flux in figure~\ref{fig:comptonflux1} on a linear scale. The blue
curve derives from the V channel and includes Coulomb correlations, whereas
the orange curve applies to the A channel. To obtain the proper flux
the curves need to be multiplied with $\CV^2$ and $5\CA^2$, respectively.

\begin{figure}[b]
\centering
\hbox to\textwidth{\includegraphics[height=5.8cm]{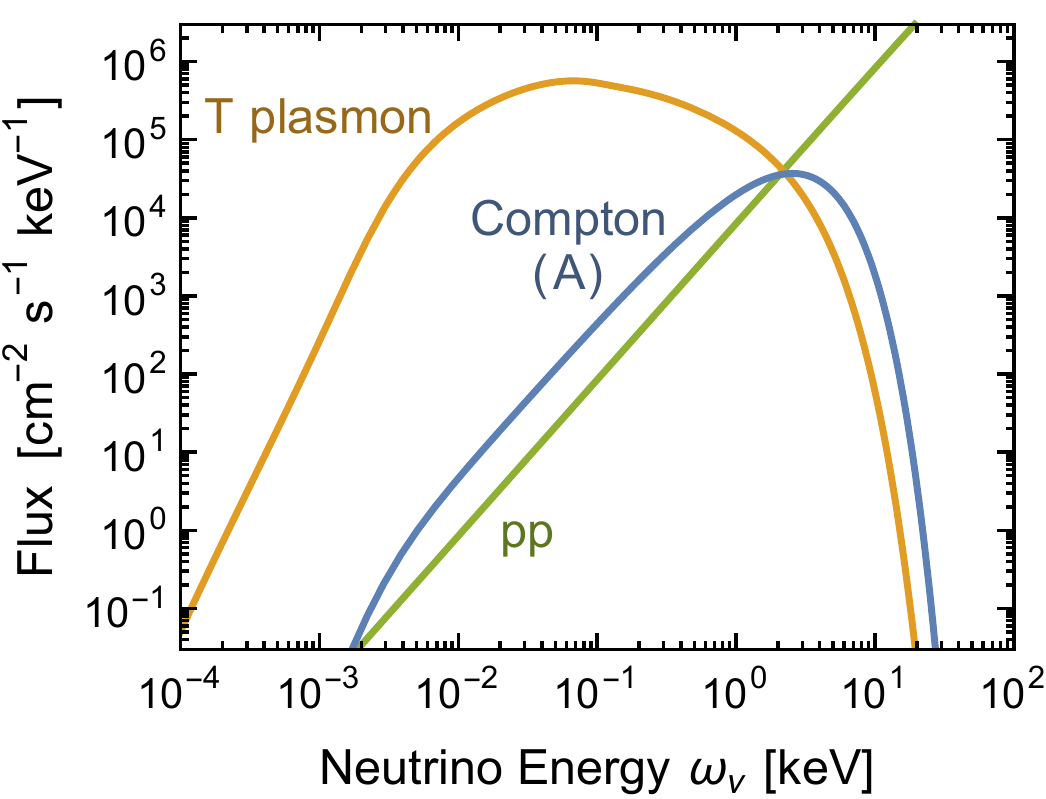}\hfil
\includegraphics[height=5.8cm]{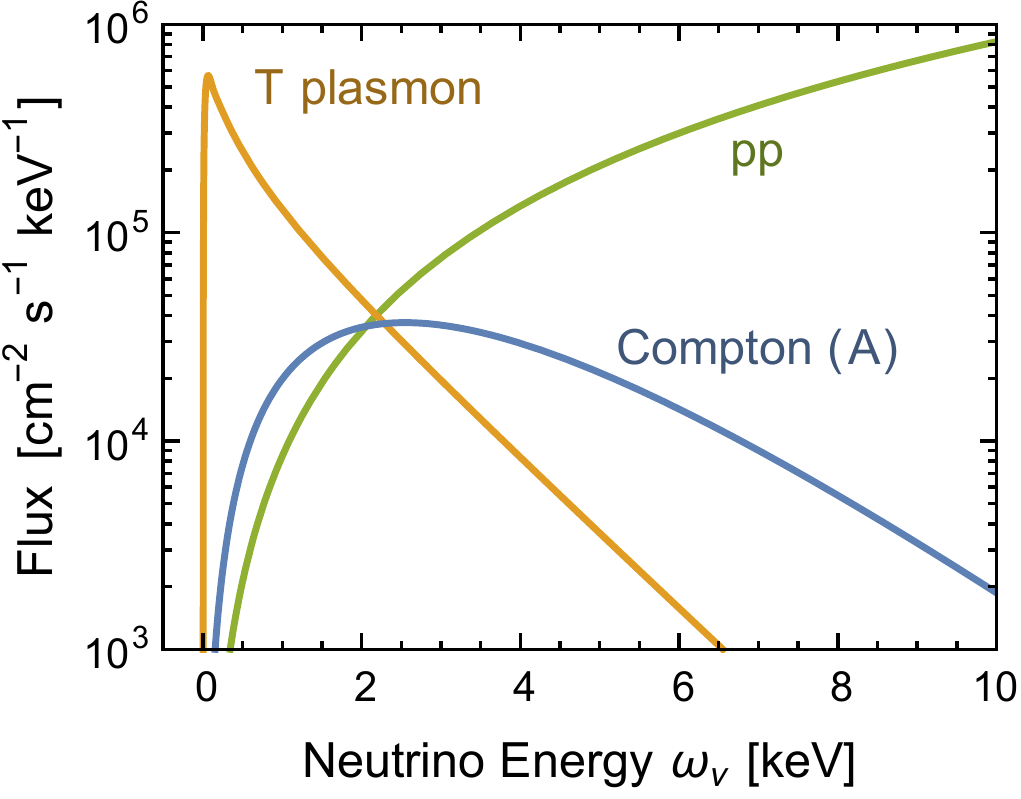}}
\caption{Solar neutrino flux at Earth from the Compton process
(axial-vector channel and only one flavor) compared with the pp flux
(only $\nu_e$) and transverse plasmon decay ($\nu_e$ and equal flux $\bar\nu_e$).
Flavor oscillations are not considered here.}
\label{fig:comptonflux2}
\end{figure}

In figure~\ref{fig:comptonflux2} we compare the axial-vector Compton flux for
a single flavor with the fluxes from T-plasmon decay and with pp neutrinos. While
we have not included the plasma frequency in the squared matrix element for the
Compton process, we do include it in the phase-space integration. In this way,
the lowest-energy Compton flux is suppressed and explains the kink in the
low-energy flux. As a consequence, the lowest-energy neutrino flux is dominated
by plasmon decay. Notice that T-plasmon decay produces almost
exclusively $\nu_e\bar\nu_e$ pairs, whereas the axial-vector Compton process produces
equal fluxes of all flavors. For $\nu_e\bar\nu_e$, there is an additional
contribution from the V channel. Apart from Coulomb-correlation corrections
and overall coefficients, the spectrum is the same as shown in
figure~\ref{fig:comptonflux1}. The flavor dependence of fluxes at Earth will
be studied later.

\section{Bremsstrahlung}
\label{sec:bremsstrahlung}

\subsection{Matrix element}

\begin{figure}[htbp]
\centering
\includegraphics[width=6cm]{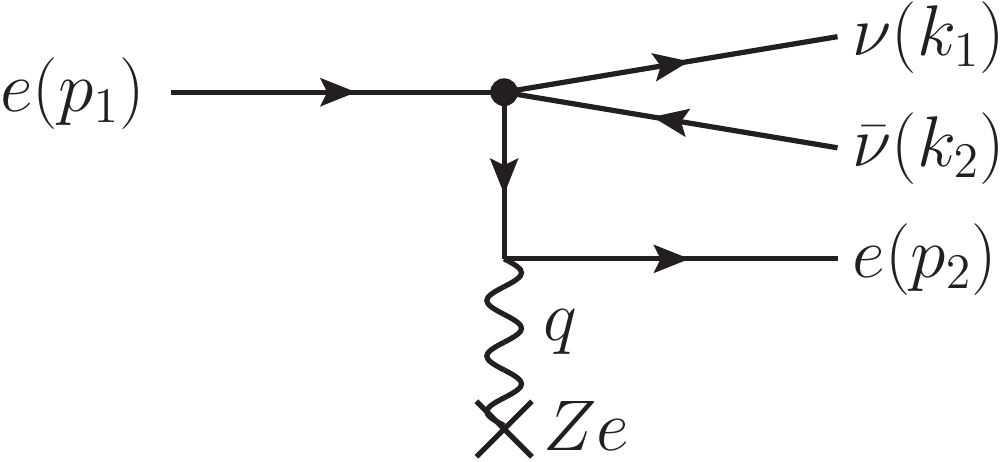}
\caption{Bremsstrahlung production of neutrino pairs. The Coulomb potential
is provided by a heavy nucleus or ion with charge $Ze$ taking up the momentum transfer
$q=(0,\bq)$. The outgoing neutrino radiation carries the four-momentum
$k=k_1+k_2=(\omega,\bk)$. There is a second diagram with the vertices exchanged.}
\label{fig:brems-graph}
\end{figure}

Next we consider bremsstrahlung production of neutrino pairs
(figure~\ref{fig:brems-graph}), also known as the free-free process,
where we consider nuclei or ions with charge $Ze$ to provide a Coulomb
potential without recoil. This process was not included in a previous
study of low-energy solar neutrino emission \cite{Haxton:2000xb}.  In
general, it is the dominant energy loss mechanism in stars with low
temperature and high electron density \cite{Cazzola:1971ru,
  Dicus:1976rj}. The first differential flux evaluation was carried
out for our nonrelativistic and nondegenerate conditions a long time
ago in reference~\cite{Gandelman:1960ex}, which has some flaws as described below,
and recently also for general conditions~\cite{Guo:2016vls}.

The scattering targets are taken to be very heavy (no recoil) with charge $Ze$ and
number density $n_Z$ and the electrons are taken be nonrelativistic. The emission rate
of neutrino pairs per unit volume and unit time is then
\begin{equation}\label{eq:brems-emission-1}
\dot n_{\nu}=n_Z \int\frac{d^3\bp_1}{(2\pi)^3}\frac{d^3\bp_2}{(2\pi)^3}
\frac{d^3\bk_1}{(2\pi)^3}\frac{d^3\bk_2}{(2\pi)^3}\,f_1(1-f_2)\,
\frac{\sum_{s_1,s_2}|\mathcal{M}|^2}{(2m_e)^2 2\omega_1 2\omega_2}\,
2\pi\delta(E_1-E_2-\omega)\,,
\end{equation}
where the sum is taken over the electron spins and $f_1$ and $f_2$ are the initial and
final-state electron occupation numbers. The final-state neutrino radiation is described by
$k=(\omega,\bk)=k_1+k_2=(\omega_1+\omega_2,\bk_1+\bk_2)$.
For the squared matrix element we find
\begin{equation}
\sum_{s_1,s_2}|\mathcal{M}|^2
=\frac{8 Z^2 e^4 }{|\bq|^4\,\omega^2}\, \(\frac{\GF^2}{2}\)
\Bigl(\CV^2 M_{\rm V}^{\mu\nu}+\CA^2 M_{\rm A}^{\mu\nu}\Bigr)N_{\mu\nu}\,,
\end{equation}
where the neutrino tensor was given in equation~(\ref{eq:neutrino-tensor})
and $\bq$ is the momentum transfer to the nucleus.

For nonrelativistic electrons, as usual we can ignore the
transfer of three-momentum to the external radiation so that
$\bq=\bp_1-\bp_2$. In this approximation we find
\begin{equation}
M_{\rm V}^{\mu\nu}=
\begin{pmatrix}
\(\frac{\bq\cdot\bk}{\omega}\)^2&&\frac{\bq\cdot\bk}{\omega}\,\bq\\[3ex]
\frac{\bq\cdot\bk}{\omega}\,\bq&&\bq^i\bq^j
\end{pmatrix}
\qquad\hbox{and}\qquad
M_{\rm A}^{\mu\nu}=
\begin{pmatrix}
\bq^2&~~&\frac{\bq\cdot\bk}{\omega}\,\bq\\[2ex]
\frac{\bq\cdot\bk}{\omega}\,\bq&~~&\(\frac{\bq\cdot\bk}{\omega}\)^2\delta^{ij}
\end{pmatrix}\,.
\end{equation}
With $\CV=\CA=1$ one should find the bremsstrahlung rates in the old literature before the discovery
of neutral currents. However, the terms proportional to $\bq{\cdot}\bk/\omega$
are missing (see the steps from equation 5 to~6 in reference \cite{Gandelman:1960ex}).
For the V-case,
bremsstrahlung arises from the electron velocity abruptly changing in a collision.
In the nonrelativistic limit, the 0-component of the electron current remains unchanged.
However, the squared matrix element is quadratic in the velocity change.
Therefore,
a consistent nonrelativistic expansion requires to go to second order in the small
velocity everywhere. If one expands the electron current only to linear order
before taking the matrix element one misses some of the terms.
A similar issue explains the factor $2/3$ difference in the axion
bremsstrahlung calculation between reference~\cite{Raffelt:1985nk} (equations 38 to 42)
and \cite{Krauss:1984gm} (equations 1 to 4).

The electron gas is assumed to be isotropic, so in equation~(\ref{eq:brems-emission-1})
we may first perform an angle average over the electron direction,
keeping their relative angle fixed. So we average over the
relative angle between $\bq$ and $\bk$, leading to
$\bq^i\bq^j\to\frac{1}{3}\bq^2\delta^{ij}$, $(\bq{\cdot}\bk)\bq\to\frac{1}{3}\,\bq^2\,\bk$,
and $(\bq\cdot\bk)^2\to\frac{1}{3}\,\bq^2\bk^2$.
Therefore, in an isotropic medium we may write
\begin{equation}
\Bigl\langle\sum_{s_1,s_2}|\mathcal{M}|^2\Bigr\rangle_{\cos (\bq,\bk)}
=\frac{8 Z^2 e^4 }{3 \bq^2}\,\(\frac{\GF^2}{2}\)\,
\frac{\(\CV^2 \bar M_{\rm V}^{\mu\nu}+\CA^2 \bar M_{\rm A}^{\mu\nu}\)N_{\mu\nu}}{\omega^4}\,,
\end{equation}
where
\begin{equation}
\bar M_{\rm V}^{\mu\nu}=
\begin{pmatrix}
\bk^2&~&\omega \bk\\[2ex]
\omega \bk&~&\omega^2\delta^{ij}
\end{pmatrix}
\qquad\hbox{and}\qquad
\bar M_{\rm A}^{\mu\nu}=
\begin{pmatrix}
3\omega^2&~&\omega \bk\\[2ex]
\omega \bk&~&\bk^2\delta^{ij}
\end{pmatrix}\,.
\end{equation}
Notice that lowering the indices in this matrix changes the sign of
the time-space part (the 0j and j0 components), i.e., lowering the indices
amounts to $\omega\bk\to-\omega\bk$. For the contractions we find
explicitly
\begin{subequations}
  \begin{eqnarray}
   \bar M_{\rm V}^{\mu\nu}N_{\mu\nu}&=&
   16\[\omega_1\omega_2(\omega_1^2+\omega_2^2+\omega_1\omega_2)
   +(\bk_1{\cdot}\bk_2)^2-\omega^2\,\bk_1{\cdot}\bk_2\]
   \nonumber\\[1ex]
   &\to&
   \frac{16}{3}\,\omega_1\omega_2\(3\omega_1^2+3\omega_2^2+4\omega_1\omega_2\),
   \\[2ex]
      \bar M_{\rm A}^{\mu\nu}N_{\mu\nu}&=&
   16\[\omega_1\omega_2(2\omega_1^2+2\omega_2^2+\omega_1\omega_2)
   -(\bk_1{\cdot}\bk_2)^2+4\omega_1\omega_2\,\bk_1{\cdot}\bk_2\]
   \nonumber\\[1ex]
   &\to&
    \frac{32}{3}\,\omega_1\omega_2\(3\omega_1^2+3\omega_2^2+\omega_1\omega_2\).
  \end{eqnarray}
\end{subequations}
The second expressions apply after an angle average over the
relative directions of $\bk_1$ and $\bk_2$ where
$\bk_1{\cdot}\bk_2\to 0$ and
$(\bk_1{\cdot}\bk_2)^2\to \frac{1}{3}\,\omega_1^2\omega_2^2$.

\subsection{Emission rate}

We may write the neutrino pair emission rate of equation~(\ref{eq:brems-emission-1})
in a way that separates the properties of the
emitted radiation (the neutrino pairs) from the properties of the medium (thermal electrons
interacting with nuclei) and find
\begin{equation}\label{eq:brems-emission-2}
\dot n_{\nu}=n_Z n_e \frac{8\,Z^2 \alpha^2}{3}
\int\frac{d^3\bk_1}{2\omega_1(2\pi)^3}\frac{d^3\bk_2}{2\omega_2(2\pi)^3}
\(\frac{\GF}{\sqrt{2}}\)^2
\frac{\(\CV^2 \bar M_{\rm V}^{\mu\nu}+\CA^2 \bar M_{\rm A}^{\mu\nu}\)N_{\mu\nu}}{\omega^4}
\,{\cal S}(\omega)\,,
\end{equation}
where $\omega=\omega_1+\omega_2$ is the energy carried away by a neutrino pair. Collecting
coefficients in a slightly arbitrary way, the relevant
response function of the medium is
\begin{equation}\label{eq:brems-response-1}
{\cal S}(\omega)=\frac{(4\pi)^2}{(2m_e)^2}\frac{1}{n_e}
\int\frac{d^3\bp_1}{(2\pi)^3}\frac{d^3\bp_2}{(2\pi)^3}\,f_1(1-f_2)\,\frac{1}{\bq^2}\,
2\pi\delta(E_1-E_2-\omega)\,,
\end{equation}
where $\bq=\bp_1-\bp_2$ is the momentum transfer in the electron-nucleus collision
mediated by the Coulomb field. We see that for both vector and axial vector emission
it is the same property of the medium causing the emission. We will
return to this point later for the free-bound and bound-bound emission
processes because we can relate both vector and axial-vector
processes to the monochromatic photon opacities, the latter providing us
essentially with ${\cal S}(\omega)$.

We now integrate over neutrino emission angles and find the
neutrino emission spectrum by using $\omega_1=\omega_\nu$ and
$\omega_2=\omega-\omega_\nu$. Integrating over the anti-neutrino
energy we find
\begin{eqnarray}\label{eq:brems-emission-3}
  \frac{d\dot n_{\nu}}{d\omega_\nu}&=&n_Z n_e\,\frac{8\,Z^2 \alpha^2}{3}\,
  \(\frac{\GF}{\sqrt{2}}\)^2\frac{1}{3\pi^4}
  \int_{\omega_\nu}^\infty d\omega\,{\cal S}(\omega)\,
  \frac{\omega_\nu^2(\omega-\omega_\nu)^2}{\omega^4}\nonumber\\[2ex]
&&\kern5em{}\times
  \Bigl[\CV^2\(3\omega^2-2\omega\omega_\nu+2\omega_\nu^2\)
  +2\CA^2\(3\omega^2-5\omega\omega_\nu+5\omega_\nu^2\)\Bigr]\,,
\end{eqnarray}
which is the rate of neutrino emission per unit volume, unit time,
and unit energy interval. The same spectrum applies to antineutrinos
because all expressions were symmetric under the exchange $1\leftrightarrow2$.

\subsection{Photon and axion emission}
\label{sec:axion-emission}

It is useful to compare the bremsstrahlung
emission rate of neutrino pairs with that of photons and axions to connect to the
previous literature and, more importantly, to relate the bremsstrahlung absorption
rate of photons to the neutrino pair emission rate.
For photon emission, the neutral-current interaction of equation~(\ref{eq:NC-interaction})
gets replaced by $i e\bar\psi_e\gamma^\mu\bar\psi_e A_\mu$. On the level of the squared
matrix element, or rather, on the level of the emission rate, this substitution
translates to
\begin{equation}\label{eq:brems-photon-emission-1}
\dot n_{\gamma}=n_Z n_e\,\frac{8\,Z^2 \alpha^2}{3}
\int\frac{d^3\bk}{2\omega(2\pi)^3}\,
e^2\,\frac{\bar M_{\rm V}^{\mu\nu}\epsilon_\mu\epsilon_\nu}{\omega^4}\,{\cal S}(\omega)\,.
\end{equation}
For the contraction we find $\bar M_{\rm V}^{\mu\nu}\epsilon_\mu\epsilon_\nu=\omega^2$.
We have used the polarization vector for a transverse photon (not a
longitudinal plasmon) so that $\bk{\cdot}{\bm \epsilon}=0$,
$\epsilon^0\epsilon^0=0$,
and ${\bm\epsilon}{\cdot}{\bm\epsilon}=1$. So the spectral photon
production rate per transverse polarization degree of freedom is
\begin{equation}\label{eq:brems-photon-emission-2}
  \frac{d\dot n_{\gamma}}{d\omega}=n_Z n_e\, \frac{8\,Z^2 \alpha^2}{3}\,
  \frac{\alpha}{\pi}\,\frac{{\cal S}(\omega)}{\omega}\,.
\end{equation}
This quantity is directly related to the medium response function
${\cal S}(\omega)$. Therefore, we can express the neutrino
emissivity of equation~(\ref{eq:brems-emission-3})
in terms of the photon emissivity as
\begin{eqnarray}\label{eq:brems-emission-4}
  \frac{d\dot n_{\nu}}{d\omega_\nu}&=&
  \frac{\GF^2}{6\pi^3\alpha}
  \int_{\omega_\nu}^\infty d\omega\,\(\frac{d\dot n_{\gamma}}{d\omega}\)\,
  \frac{\omega_\nu^2(\omega-\omega_\nu)^2}{\omega^3}\nonumber\\[2ex]
&&\kern4em{}\times
  \Bigl[\CV^2\(3\omega^2-2\omega\omega_\nu+2\omega_\nu^2\)
  +2\CA^2\(3\omega^2-5\omega\omega_\nu+5\omega_\nu^2\)\Bigr]\,.
\end{eqnarray}
Therefore, given the spectral photon emissivity, e.g.\ taken from the
photon opacity calculation, we can directly extract the neutrino
emission spectrum. We will return to this point in section~\ref{sec:freebound}.

Axions couple to the electron axial current with an interaction of the
derivative form
$(C_e/2f_a)\,\bar\psi_e\gamma^\mu\gamma_5\psi_e\,\partial_\mu a$,
where $a$ is the axion field, $f_a$ the axion decay constant,
and $C_e$ a model-dependent numerical coefficient. One often
uses a dimensionless axion-electron Yukawa coupling
$g_{ae}=C_e m_e/f_a$ so that the interaction is
$(g_{ae}/2m_e)\,\bar\psi_e\gamma^\mu\gamma_5\psi_e\,\partial_\mu a$.
The bremsstrahlung emission rate is found to be
\begin{equation}\label{eq:brems-axion-emission-1}
\dot n_{a}=n_Z n_e\, \frac{8\,Z^2 \alpha^2}{3}
\int\frac{d^3\bk}{2\omega(2\pi)^3}\,
\(\frac{g_{ae}}{2m_e}\)^2
\frac{\bar M_{\rm A}^{\mu\nu}k_\mu k_\nu}{\omega^4}\,{\cal S}(\omega)\,.
\end{equation}
For massless axions with $\omega=|\bk|$ we find for the contraction
$\bar M_{\rm A}^{\mu\nu}k_\mu k_\nu=2\omega^4$. Therefore, the
spectral emissivity is
\begin{equation}\label{eq:brems-axion-emission-2}
  \frac{d\dot n_{a}}{d\omega}=n_Z n_e\,\frac{8\,Z^2 \alpha^2}{3}\,
  \(\frac{g_{ae}}{2m_e}\)^2
    \frac{\omega}{2\pi^2}\,{\cal S}(\omega)\,.
\end{equation}
Notice that this spectrum is harder than the photon spectrum by a
factor $\omega^2$ caused by the derivative structure of the axion
interaction. Still, fundamentally it depends on the same
medium response function ${\cal S}(\omega)$.
We have checked that in the nondegenerate limit this axion
emission rate agrees with reference~\cite{Raffelt:1985nk}.

\subsection{Medium response function and screening effects}

The medium response function defined in
equation~(\ref{eq:brems-response-1}) could be easily evaluated if the
nuclei used as scattering centers were uncorrelated. However, their
Coulomb interaction leads to anticorrelations encoded in an ion
structure factor $S_i(\bq^2)$ similar to the case of electron-electron
correlations discussed in section~\ref{sec:correlatoins}.  Therefore,
under the integral in equation~(\ref{eq:brems-response-1}) we need to
include $S_i(\bq^2)$, which we will discuss later. We mention in
passing that ${\cal S(\omega)}$ given in
equation~(\ref{eq:brems-response-1}), with or without including
$S_i(\bq^2)$, fulfills the detailed-balancing condition
${\cal S}(-\omega)={\cal S}(\omega)\,e^{\omega/T}$. Here a negative $\omega$ means
energy absorbed by the medium, whereas a positive $\omega$ for us
always means energy emitted, although in the literature one usually uses
the opposite convention.

To write ${\cal S}(\omega)$ in a more compact form we first note that the
electron number density is given by
equation~(\ref{eq:electron-density}) in terms of the nonrelativistic
degeneracy parameter $\eta$. We further write the kinetic electron
energy in dimensionless form as $u=\bp^2/(2m_e T)$ so that the
occupation number is $f_u=1/(e^{u-\eta}+1)$. Then the structure
function is
\begin{equation}
  {\cal S}(\omega)=\frac{\pi}{m_e\sqrt{2m_eT}}\,s(\omega/T)\,
\end{equation}
where
\begin{equation}
  s(w)=
  \int_0^\infty\!du_2\,\frac{\sqrt{u_1 u_2}}{e^{u_1-\eta}+1}\,
  \frac{1}{e^{-(u_2-\eta)}+1}
\int_{-1}^{+1}\!d{\rm c}_\theta\,\frac{2m_eT}{\bq^2}\,S_i(\bq^2)
\bigg/\int_0^\infty du\,\frac{\sqrt{u}}{e^{u-\eta}+1}\,.
\end{equation}
Here $u_1=u_2+w$ and $w=\omega/T$. Moreover,
$\bq^2=|\bp_1-\bp_2|^2=\bp_1^2+\bp_2^2-2|\bp_1||\bp_2|{\rm c}_\theta$
with ${\rm c}_\theta=\cos\theta$. Therefore,
$\bq^2/(2m_eT)=u_1+u_2-2\sqrt{u_1u_2}\,{\rm c}_\theta$.

\begin{figure}[ht]
\centering
\includegraphics[width=7cm]{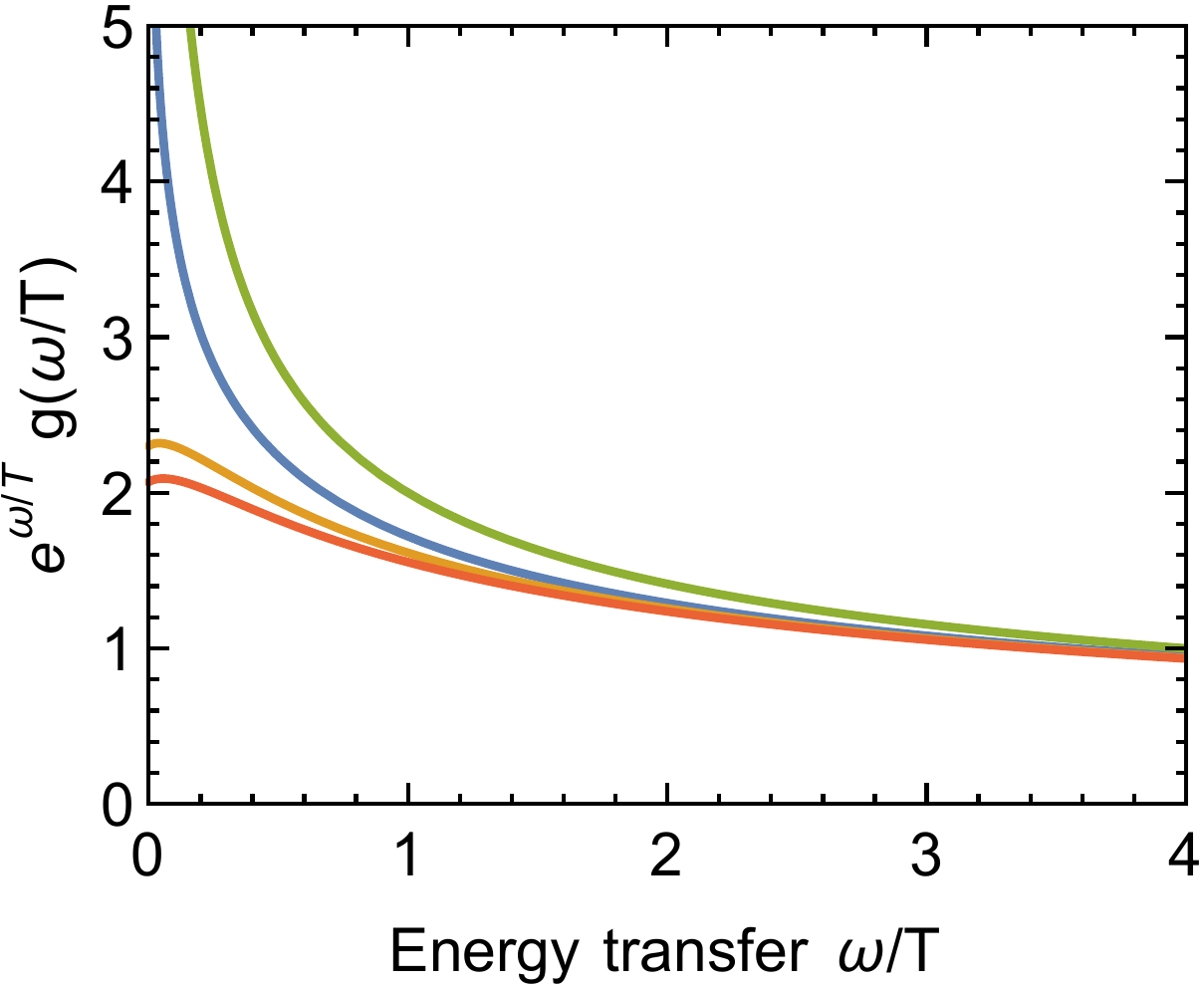}
\caption{Structure function $e^w s(w)$ for bremsstrahlung.
{\em Blue line:} Non-degenerate electrons
given in equation~(\ref{eq:sND}) without Coulomb correlations ($\mu=0$).
{\em Orange line:} Coulomb correlations included with $\mu=0.04$
appropriate for the solar center.
{\em Red line:} Electron degeneracy included with $\eta=-1.4$ appropriate
for the solar center.
{\em Green line:} Asymptotic form for large $w=\omega/T$.}
\label{fig:sND}
\end{figure}

Besides the original squared matrix element, this function depends on
electron degeneracy effects and
Coulomb correlation effects encoded in $S_i(\bq^2)$. Anticipating that electron degeneracy is
not a large correction we first reduce this expression to Maxwell-Boltzmann rather than
Fermi-Dirac statistics. Formally this is the $\eta\to-\infty$ limit, leading to the
non-degenerate (ND) structure function
\begin{equation}
  s_{\rm ND}(w)=
  \int_0^\infty\!du\,e^{-u-w}\sqrt{(u+w)\,u}\,
  F_i(u,w)
\bigg/\int_0^\infty du\,e^{-u}\sqrt{u}\,,
\end{equation}
where the integral in the denominator is simply $\sqrt{\pi}/2$.
The integral kernel is
\begin{equation}
F_i(u,w)=\int_{-1}^{+1}\!d{\rm c}_\theta\,\frac{2m_eT\,S_i(\bq^2)}{\bq^2}\,.
\end{equation}
The ion structure function from Coulomb correlation effects will be an expression of the type given in equation~(\ref{eq:S-Coulomb}), but with the role of electrons and ions interchanged. However, in a multi-component plasma, an exact treatment is difficult; because screening will be a relatively small correction, we use
\begin{equation}\label{eq:screeningbrem}
S_i(\bq^2)=\frac{\bq^2}{\bq^2+k_{\rm s}^2}\,,
\end{equation}
where $k_{\rm s}$ is a phenomenological screening scale. We use $k_i$ given in equation~(\ref{eq:screening-scales}) for the ion correlations.
With $\mu=k_{\rm s}^2/(2m_eT)$ we therefore use
\begin{eqnarray}
F_i(u,w)&=&\int_{-1}^{+1}\!d{\rm c}_\theta\,\frac{1}{\mu+2u+w-2\sqrt{(u+w)\,u}\,{\rm c}_\theta}
\nonumber\\[2ex]
&=&\frac{1}{2\sqrt{(u+w)\,u}}\log\(\frac{\mu+2u+w+2\sqrt{(u+w)\,u}}{\mu+2u+w-2\sqrt{(u+w)\,u}}\)\,.
\end{eqnarray}
Without screening ($\mu=0$) this expression diverges logarithmically for small $w$. However, for axion emission and neutrino pair emission, this divergence is moderated by at least one power of $\omega$, so even without correlation effects, the emission of soft radiation does not diverge. Near the solar center one finds $k_i=7~{\rm keV}$ and with $T=1.3~{\rm keV}$ one finds $\mu=0.037\ll 1$, so screening is not a strong effect. Overall then the ND structure function is
\begin{equation}\label{eq:sND}
  s_{\rm ND}(w)=\frac{e^{-w}}{\sqrt{\pi}}
  \int_0^\infty\!du\,e^{-u}\,\log\(\frac{\mu+2u+w+2\sqrt{(u+w)\,u}}{\mu+2u+w-2\sqrt{(u+w)\,u}}\)
  \to\frac{2e^{-w}}{\sqrt{w}}\quad(\hbox{large $w$})
\end{equation}
In figure~\ref{fig:sND} we show $e^w s_{\rm ND}(w)$ with and without Coulomb correlation effects
and the asymptotic form for large $w$.
We also show as a red line $e^w s(w)$ including Fermi-Dirac statistics for the electrons with
$\eta=-1.4$, appropriate for the solar center.

While correlation effects strongly modify the structure function at low energy transfer, this effect is much smaller after folding with the neutrino phase space according to equation~\eqref{eq:brems-emission-3}. Even without correlations, the neutrino spectrum does not diverge at low energies. Including Coulomb correlations reduces it at low energies by some 20\% and only by very little in the main keV-range of the spectrum. Pauli blocking of final states provides a further 5\% suppression effect at very low energies, so overall these are fairly minor effects.

\subsection{Electron-electron bremsstrahlung}
\label{sec:ebremsstrahlung}

The electron-electron bremsstrahlung process is similar to
electron-proton bremsstrahlung with a number of important
modifications \cite{Raffelt:1985nk}. The vector-current emission rate
is of higher order in velocity---the simple dipole term vanishes in
the scattering of equal-mass particles. In the non-degenerate and
non-relativistic limit and ignoring Coulomb correlations, the
axial-vector rate is $1/\sqrt2$ that of the electron-proton rate.  In
other words, we obtain the $ee$ bremsstrahlung rate from the
axial-current part of equation~\eqref{eq:brems-emission-3} with the
substitution $Z^2 n_Z n_e\to n_e^2/\sqrt{2}$.

Taking degeneracy effects and Coulomb correlations exactly into
account would be very hard.  Instead we simply add the
electron-electron term to the electron-ion one and therefore use the
same treatment as for the latter. These corrections are rather small
and $ee$ bremsstrahlung is subdominant, so the overall error
introduced by this approach is on the level of a few percent.

\subsection{Solar neutrino flux}

\begin{figure}[b!]
\centering
\includegraphics[width=7.2cm]{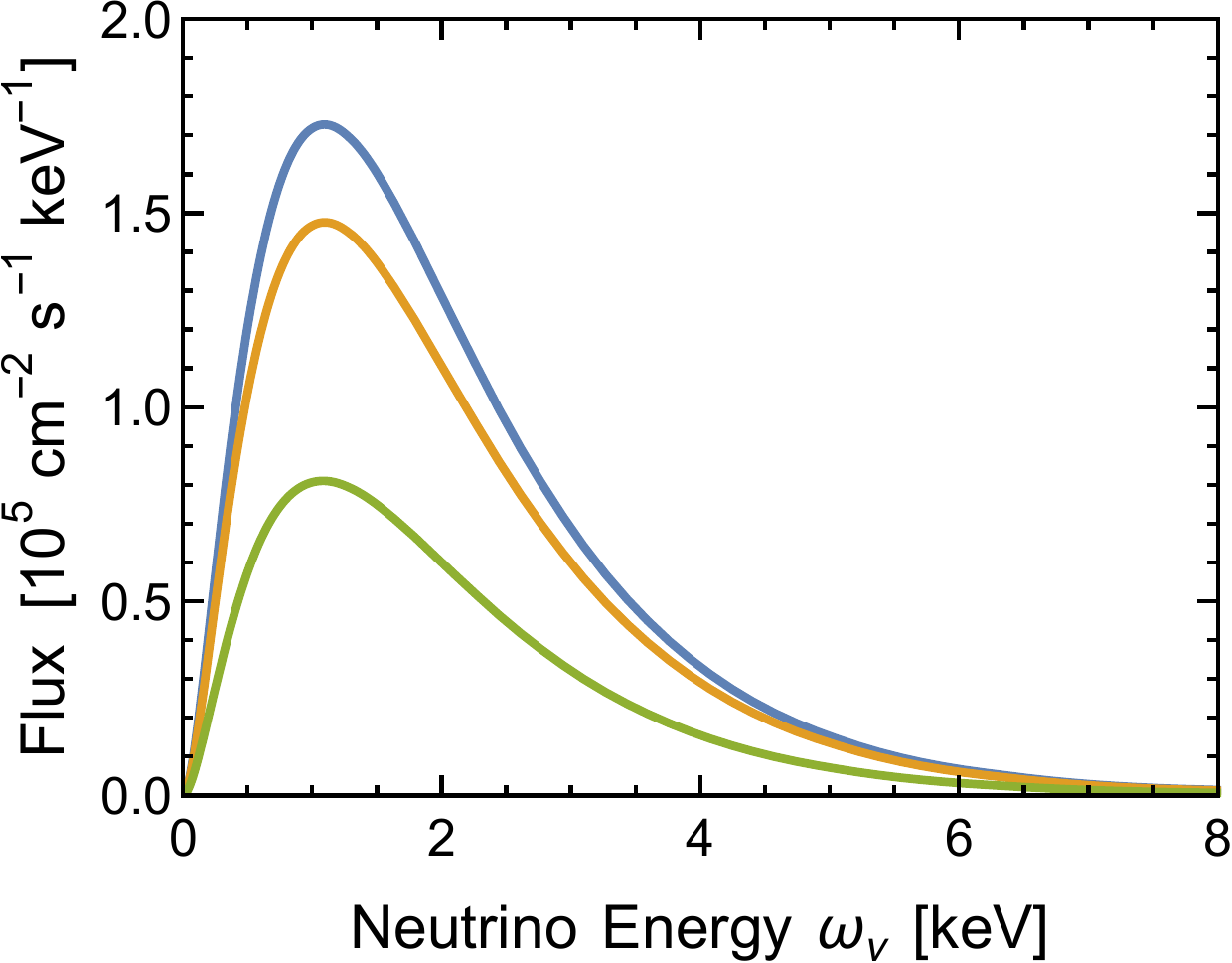}
\caption{Neutrino flux from bremsstrahlung production for the vector
  (blue) and axial-vector (orange) electron-ion interaction, and from the $ee$ interaction (green) which contributes only in the
  axial-vector channel. For the proper flux, the blue
  curve is to be multiplied with $\CV^2$, the orange and green
  curves with $2\CA^2$.}
\label{fig:brems1}
\end{figure}

As a final step we integrate the bremsstrahlung emission rate over our
standard solar model to obtain the neutrino flux at Earth. In
figure~\ref{fig:brems1} we show separately the vector and axial-vector
contributions from electron-ion scattering as well as the one from
electron-electron scattering which only contributes in the axial
channel. These curves need to be multiplied with the flavor-dependent
values of $\CV^2$ and $2\CA^2$ to obtain the proper fluxes. For the
electron-ion contributions, we include only hydrogen and helium as
targets. The contribution from metals is only a few percent and
will be included in the opacity-derived flux in
section~\ref{sec:freebound}.

\begin{figure}[htbp]
\centering
\hbox to\textwidth{\includegraphics[height=5.8cm]{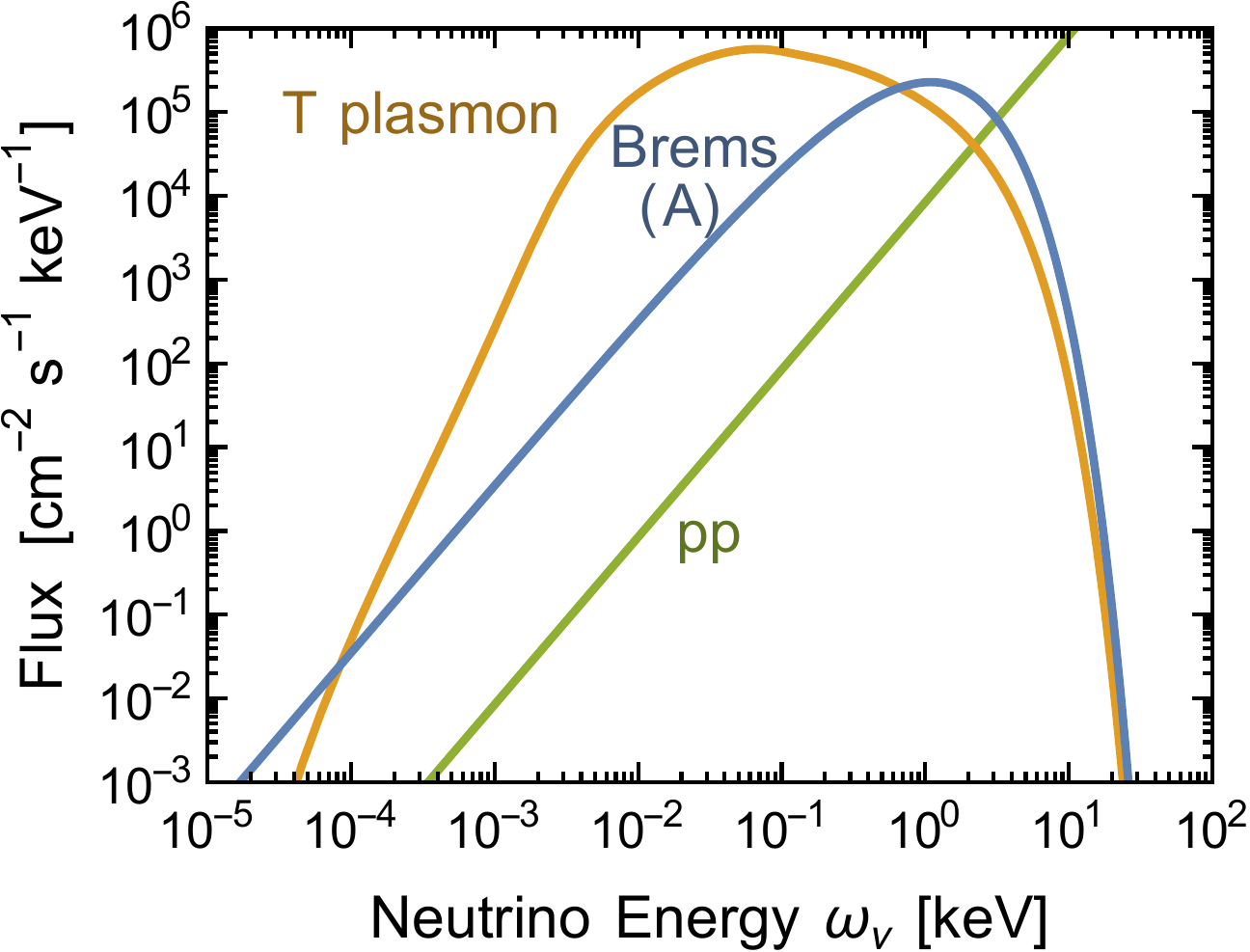}\hfil
\includegraphics[height=5.8cm]{pics/fig14a.pdf}}
\caption{Solar neutrino flux at Earth from the axial-channel
  bremsstrahlung process ($eI$ and $ee$ contributions) for one
  flavor, compared with the pp flux (only $\nu_e$) and transverse
  plasmon decay ($\nu_e$ and equal flux $\bar\nu_e$).  Flavor
  oscillations are not considered here.}
\label{fig:brems2}
\end{figure}

Finally we show in figure~\ref{fig:brems2} the axial-channel
bremsstrahlung flux for one flavor in comparison with the $\nu_e$ flux
from the nuclear pp reaction and from T plasmon decay. Similar to the
Compton process, the bremsstrahlung flux becomes important in the
cross-over region between the T-plasmon and pp fluxes.

Bremsstrahlung is the dominant contribution at very low energies.
From equation~\eqref{eq:brems-emission-3} we see that for very low neutrino energies, the spectrum varies as $\omega_\nu^2$, independently of details of the structure function ${\cal S}(\omega)$. This scaling remains true after integrating over the Sun, so the very-low energy thermal emission spectrum scales as $\omega_\nu^2$ and thus in the same way as the pp flux given in equation~\eqref{eq:ppflux}.

\subsection{Beyond the Born approximation}

Traditionally the bremsstrahlung emission rate of neutrinos and other particles is
calculated in Born approximation starting with the usual Feynman rules. However, the
bremsstrahlung emission by a non-relativistic electron scattering on an ion receives a significant modification if one uses appropriately modified electron wave functions instead of plane waves,
a point first discussed in the context of x-ray production in free-free transitions a long time ago by Sommerfeld \cite{Sommerfeld:1931}. Such an enhancement in bremsstrahlung emission arises because of the long-range Coulomb potential. It is the counterpart of what is known as Fermi-Coulomb function in the context of beta decay and of the so-called Sommerfeld enhancement, to be taken into account in dark matter annihilation processes \cite{ArkaniHamed:2008qn}. In these cases, the correction is simply given by $f(v)=|\psi(0)|^2$, i.e., by the normalization of the outgoing (or ingoing) Coulomb distorted wave function
\begin{equation}
\sigma=\sigma_0 f(v)=\sigma_0 \frac{2\pi Z \alpha}{v}\frac{1}{1-e^{\frac{2\pi Z\alpha}{v}}}\,,
\end{equation}
where $\sigma_0$ is the cross section evaluated with plane waves and $v$ is the velocity of the outgoing (or ingoing) particle.

\begin{figure}[b!]
\centering
\hbox to\textwidth{\includegraphics[scale=0.72]{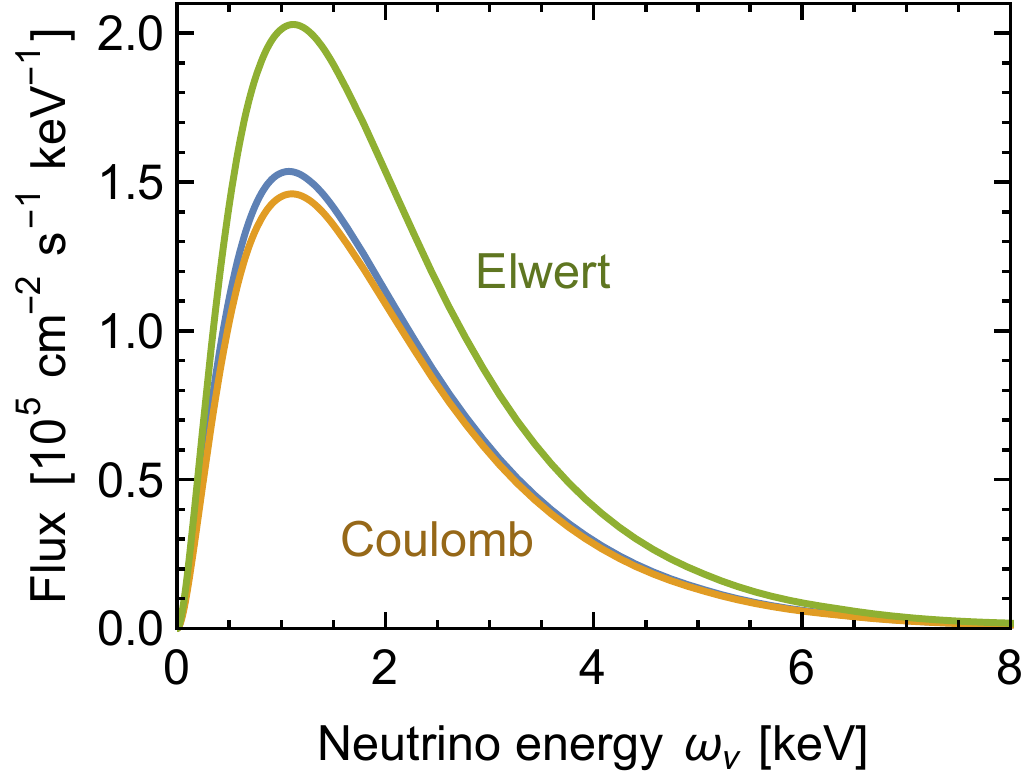}\hfill
\includegraphics[scale=0.72]{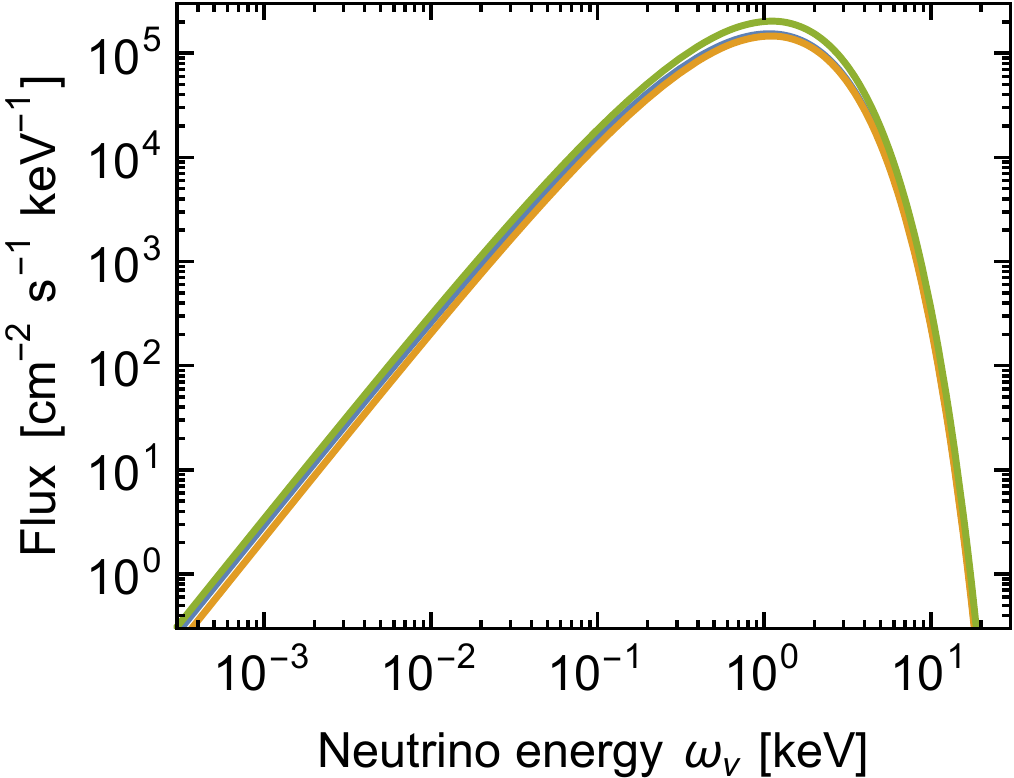}}
\vskip12pt
\hbox to\textwidth{\includegraphics[scale=0.72]{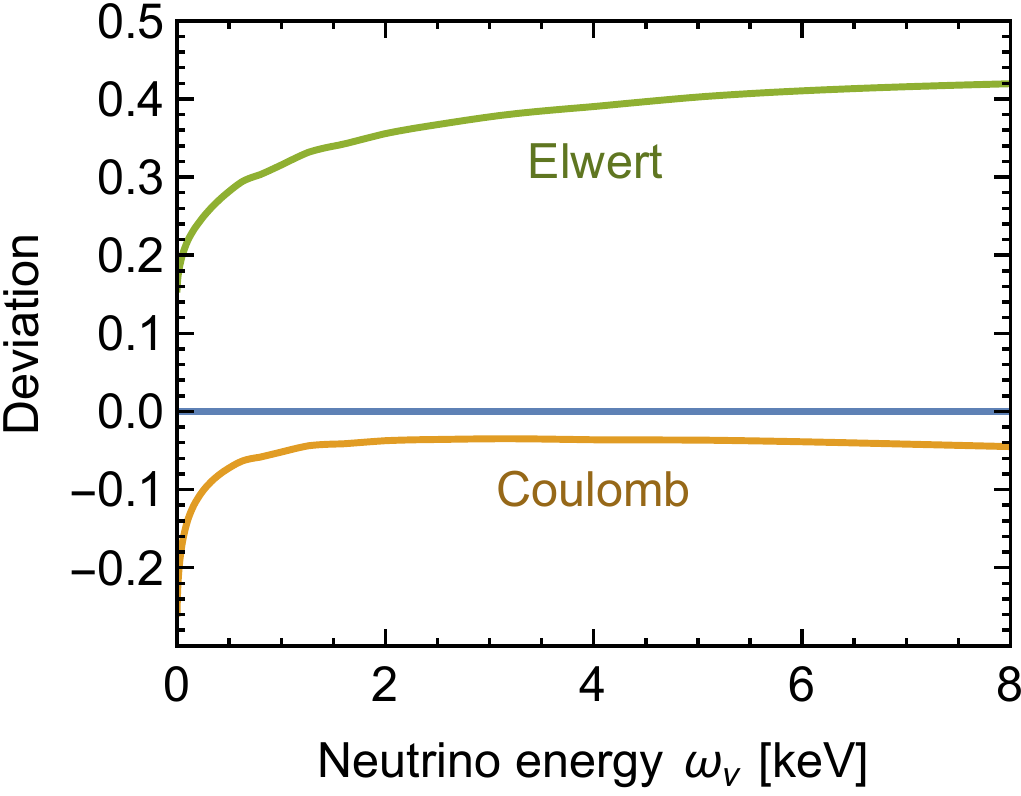}\hfill
\includegraphics[scale=0.72]{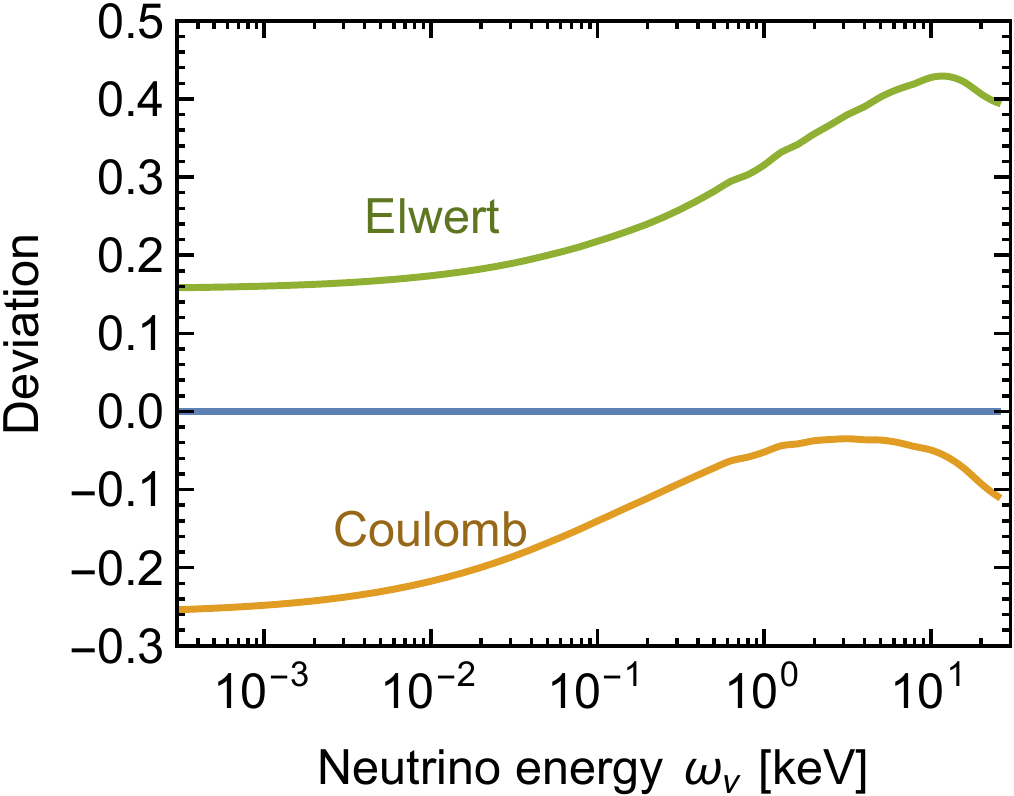}}
\caption{Solar neutrino flux from bremsstrahlung production for axial-vector electron-ion interaction, without corrections (blue), and including respectively correlations (orange) and the Elwert factor (green). For the proper flux, the
  curves in the upper panels are to be multiplied with $2\CA^2$.}
\label{fig:bremssgaunt}
\end{figure}

Such corrections have been extensively studied also for bremsstrahlung (see e.g.\ reference~\cite{Karzas:1961}). Elwert found that a good approximation to correct the Born scattering formula is obtained by simply multiplying equation~(\ref{eq:brems-response-1}) by the factor \cite{Elwert:1939}
\begin{equation}
f_E=\frac{v_{\rm i}}{v_{\rm f}}\frac{1-e^{\frac{2\pi Z \alpha}{v_{\rm i}}}}{1-e^{\frac{2\pi Z \alpha}{v_{\rm f}}}} ,
\end{equation}
which is the ratio $|\psi_{\rm f}(0)|^2/|\psi_{\rm i}(0)|^2$ of the final and initial state electron wave functions. In figure~\ref{fig:bremssgaunt} (green line) we show the effect on the overall solar neutrino flux of including this factor in the bremsstrahlung rate, leading to a typical 20--30\% enhancement. We also show as an orange line the effect of including Coulomb correlations which reduce the flux typically by some 5\%. At very small energies, the Elwert factor becomes less important and Coulomb correlations more important.

As noted by one of us in the context of solar axion emission~\cite{Redondo:2013wwa}, the Sommerfeld correction is included in the photon opacity calculation. On the other hand, using unscreened Coulomb wave functions in a stellar plasma is not fully consistent---the true correction should be considerably smaller, especially for bremsstrahlung on hydrogen and helium. Therefore, as in reference~\cite{Redondo:2013wwa} we calculate these rates directly, not from the opacities, and leave out the Elwert factor, possibly underestimating the true flux by some 10\%. On the other hand, we will include the Coulomb correlation factor. At keV-range energies this makes no big difference, but should be a reasonable correction in the far sub-keV range where bremsstrahlung is the dominant flux
and the Elwert factor is small.

\section{Free-bound and bound-bound transitions}
\label{sec:freebound}

\begin{figure}[b!]
\centering
\includegraphics[height=5.8cm]{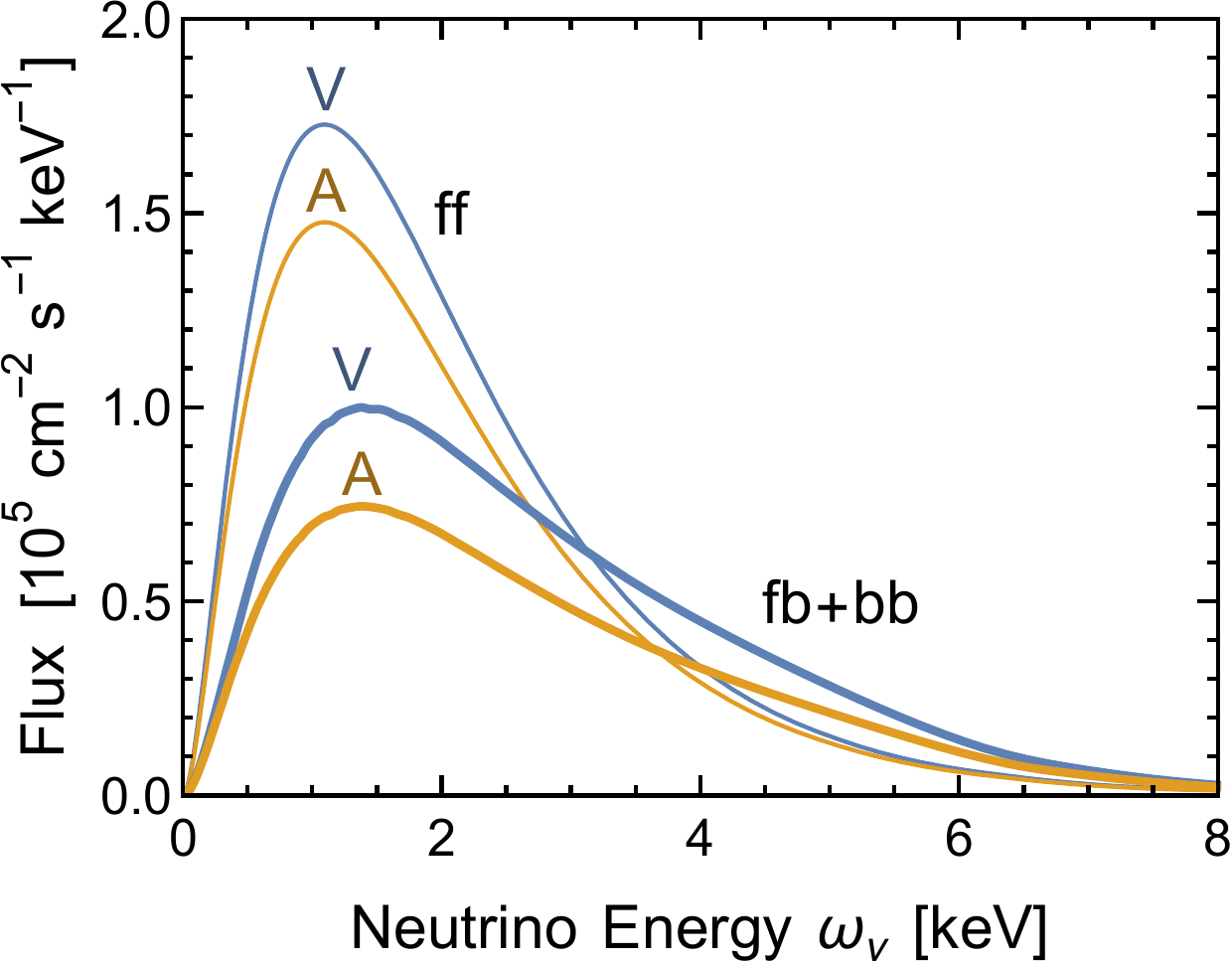}
\caption{Solar neutrino flux at Earth from free-free (ff), free-bound (fb) and
  bound-bound (bb) electron-ion transitions for the vector (V) and axial-vector (A)
  contributions. The proper fluxes are found by multiplying the curves
  with $\CV^2$ and $2\CA^2$, respectively. The ff curves
  include bremsstrahlung on hydrogen and helium and are the same as in
  figure~\ref{fig:brems1}; they exclude
  electron-electron bremsstrahlung. The fb+bb curves include
  bremsstrahlung on metals.}
\label{fig:fb-bb}
\end{figure}

The nuclei of the solar plasma are imperfectly ionized, notably the
``metals'' (elements heavier than helium). Therefore, in addition to
bremsstrahlung (free-free electron transitions), free-bound (fb) and
bound-bound (bb) transitions are also important for particle
emission. In the context of axion emission by electrons, these
processes imprint a distinctive line pattern on the expected solar
axion flux \cite{Redondo:2013wwa}. Likewise, the photon opacities, as
input to solar models, depend strongly on these processes. In
reference~\cite{Haxton:2000xb} free-bound processes were included by
explicit atomic transition calculations for a number of elements.

However, following the approach taken by one of us in an earlier
paper for calculating the solar axion flux \cite{Redondo:2013wwa}, the
emission rate can be related to the photon opacity
by equation~\eqref{eq:brems-emission-4}, i.e., the neutrino emissivity
is the same as the phase-space weighted photon
emissivity. Therefore, one can use the solar photon opacity from the
literature to derive the neutrino emissivity. In
section~\ref{sec:axion-emission} we have derived the relation between photon and neutrino emissivity explicitly for bremsstrahlung, but it
applies in this form to all processes where a nonrelativistic
electron makes a transition in the potential of an external scattering center
which takes up momentum. In other words, it applies in the long-wavelength
approximation with regard to the electron.  On the other hand, this relation does not apply to electron-electron bremsstrahlung, the Compton process, or plasmon decay.

While we could have used this relation to extract the bremsstrahlung
emissivity from the opacities, we have preferred to treat
bremsstrahlung on hydrogen and helium as well as electron-electron
bremsstrahlung explicitly in the interest of completeness. For the fb and bb transitions as well as
bremsstrahlung on metals we proceed as in
reference~\cite{Redondo:2013wwa} to extract the neutrino
emissivity. In figure~\ref{fig:fb-bb} we show the resulting neutrino
flux spectrum at Earth in comparison with the bremsstrahlung result on
hydrogen and helium derived earlier. We show the vector and
axial-vector contributions, each to be multiplied with the
flavor-dependent $\CV^2$ or $2\CA^2$ to arrive at the proper flux. While the
fb and bb contributions are subdominant relative to
bremsstrahlung, they dominate at higher energies. This behavior
was to be expected because
they are more relevant than ff processes in the Rosseland opacities at
any radius (see e.g.\ figure~15 of reference~\cite{Krief:2016znd}).

\section{Solar neutrino flux at Earth}
\label{sec:solarflux}

\subsection{Flavor-dependent fluxes}

\begin{figure}[htbp]
\centering
\hbox to\textwidth{\includegraphics[height=6.9cm]{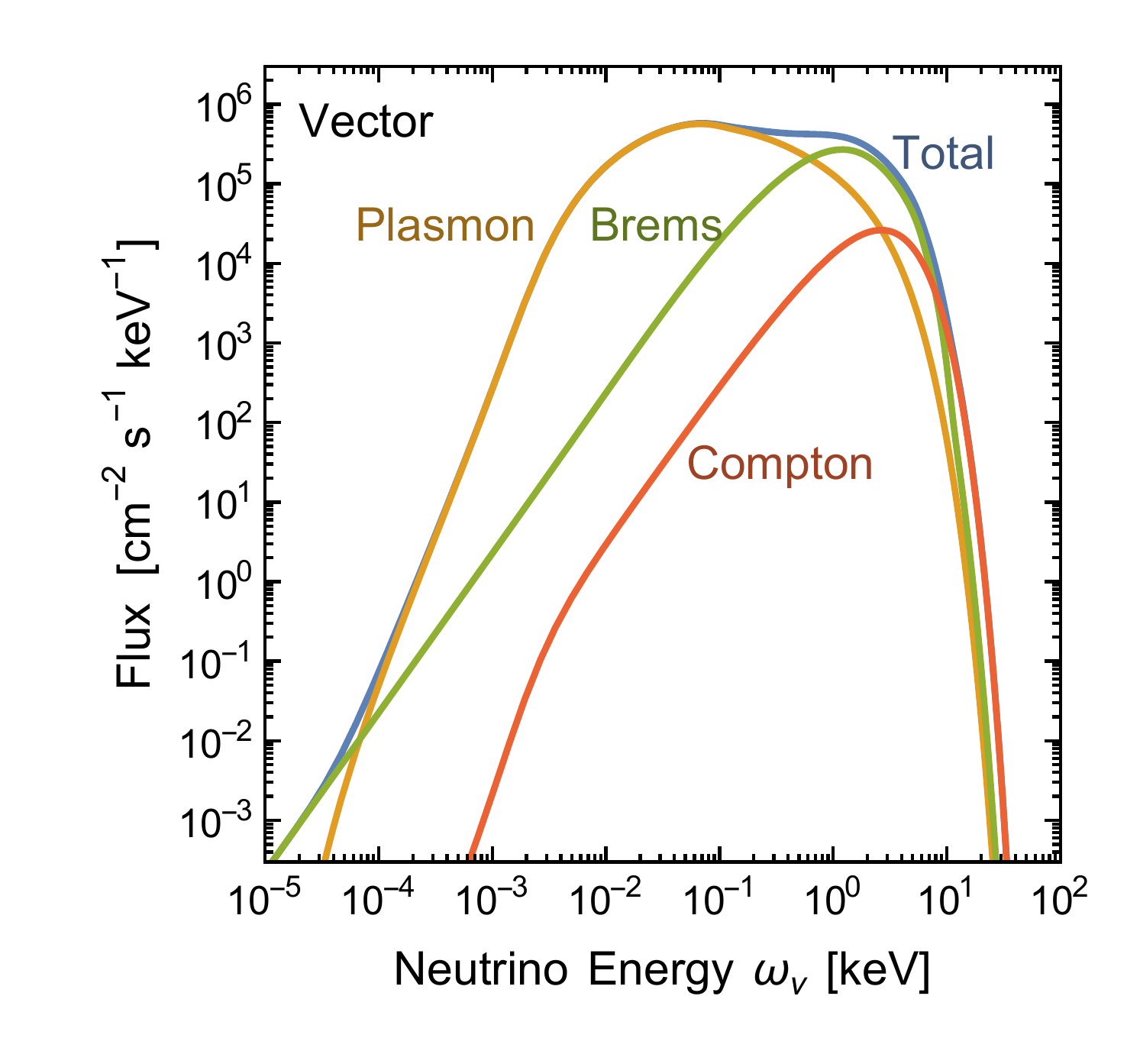}
\hfil\includegraphics[height=6.9cm]{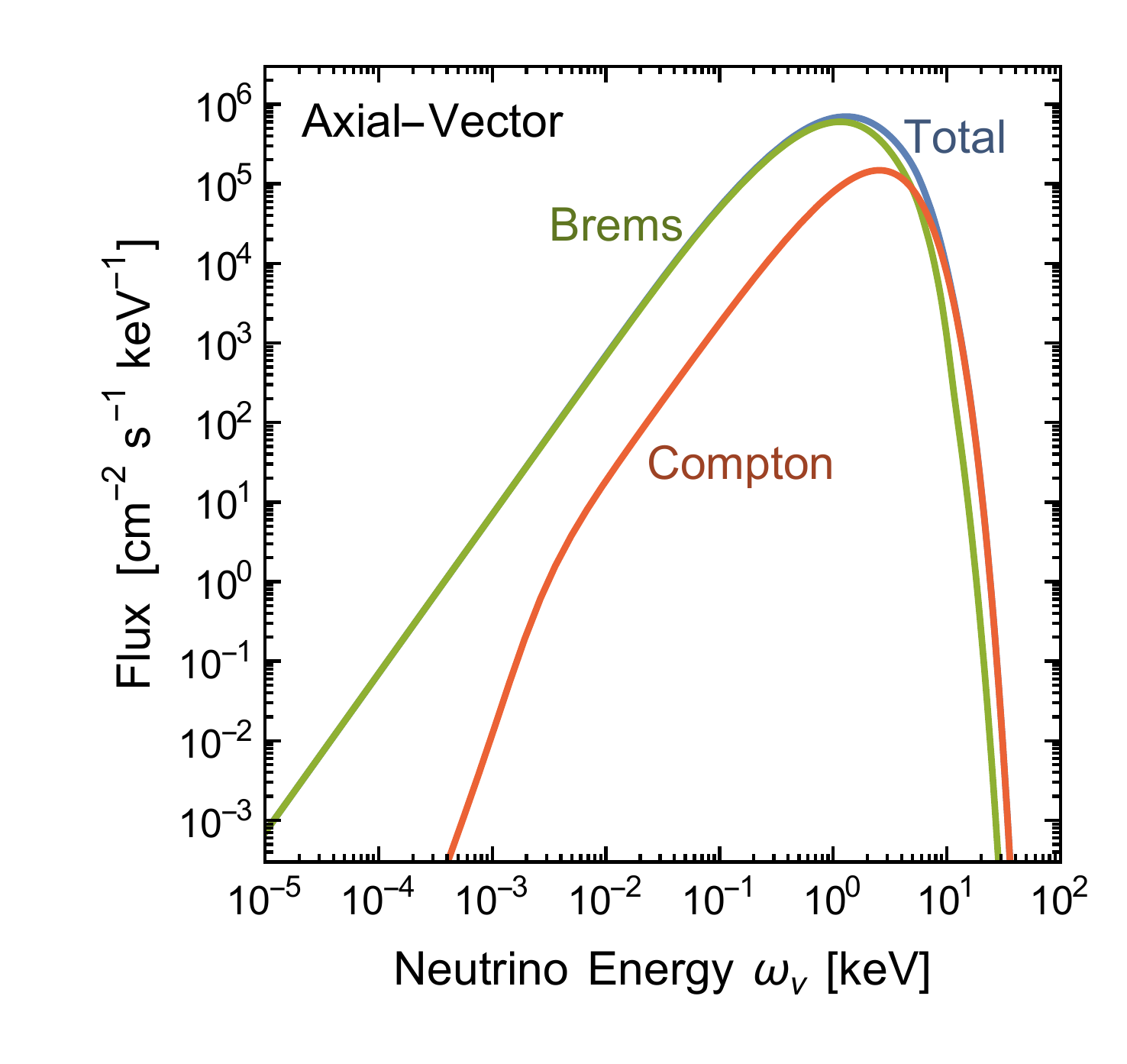}}
\caption{Solar neutrino flux at Earth from the indicated
  processes, where ``brems'' includes ff, fb and bb.
  {\em Left panel:} Vector coupling, for proper
  flux multiply with $\CV^2$. {\em Right panel:} Axial-vector coupling, for proper
  flux multiply with $\CA^2$.}
\label{fig:V-A-flux}
\end{figure}

In order to consolidate the solar flux results from different
thermal processes we show the spectra at Earth in
figure~\ref{fig:V-A-flux}. In the left panel we show all contributions
relevant for the vector coupling, where the true flux is found by
multiplication with the flavor-dependent value of $\CV^2$. At low
energies, plasmon decay dominates, at intermediate ones
bremsstrahlung, and at the highest energies Compton process.
In the right panel we show the analogous axial-vector result which does not
have any significant plasmon-decay contribution.

The intrinsic uncertainties of the various emissivity calculations
should not be larger than a few tens of percent concerning various
issues of in-medium effects such as correlations or Coulomb wave
functions of charged particles. In addition, every process has a
different dependence on temperature, density and chemical composition
so that different standard solar models will produce somewhat
different relative weights of the different processes. We have not
studied the variation of the different flux contributions depending on
different standard solar models, but the overall uncertainty again
should be in the general ten percent range.

\begin{figure}[htbp]
\centering
\hbox to\textwidth{\includegraphics[height=5.8cm]{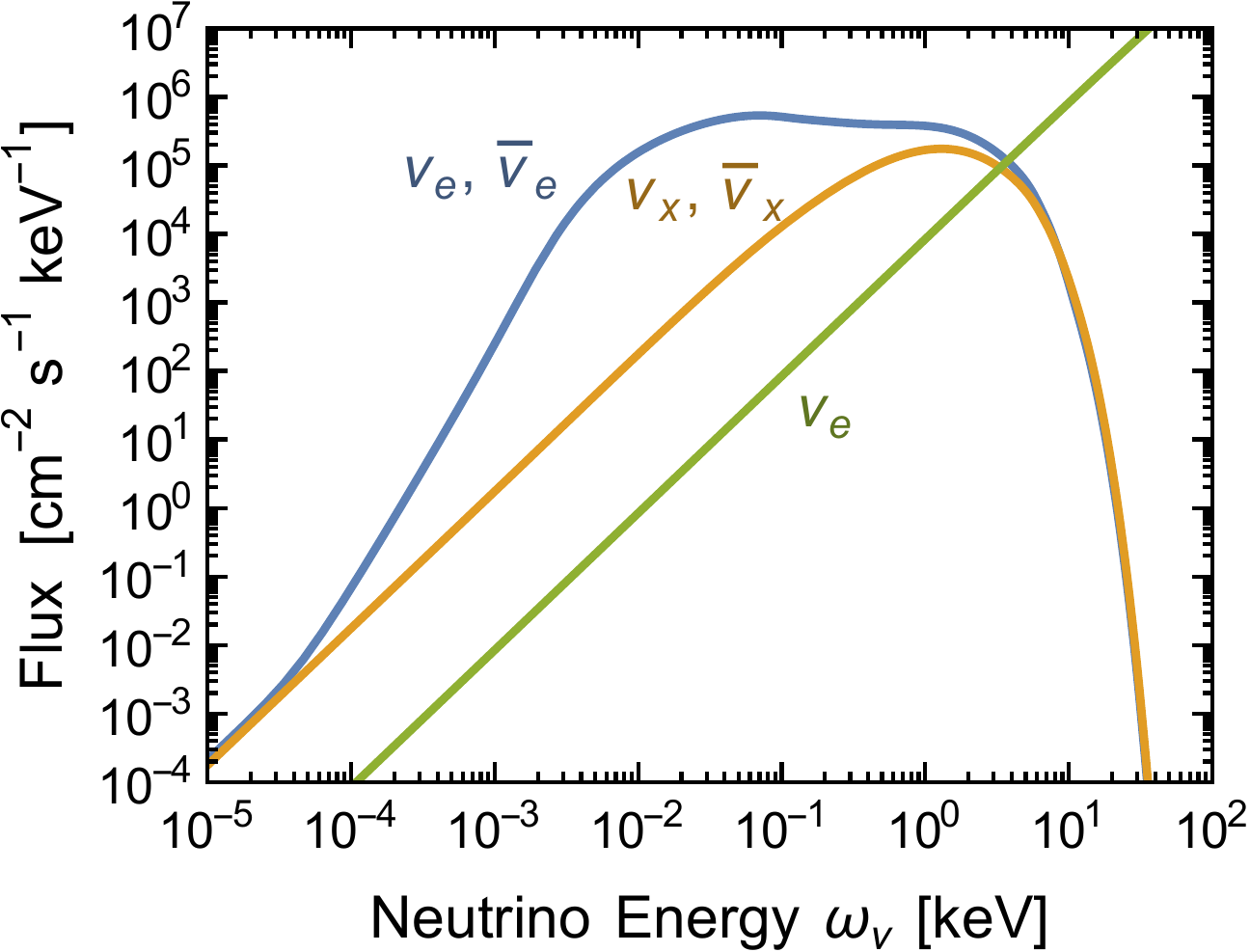}
\hfil\includegraphics[height=5.8cm]{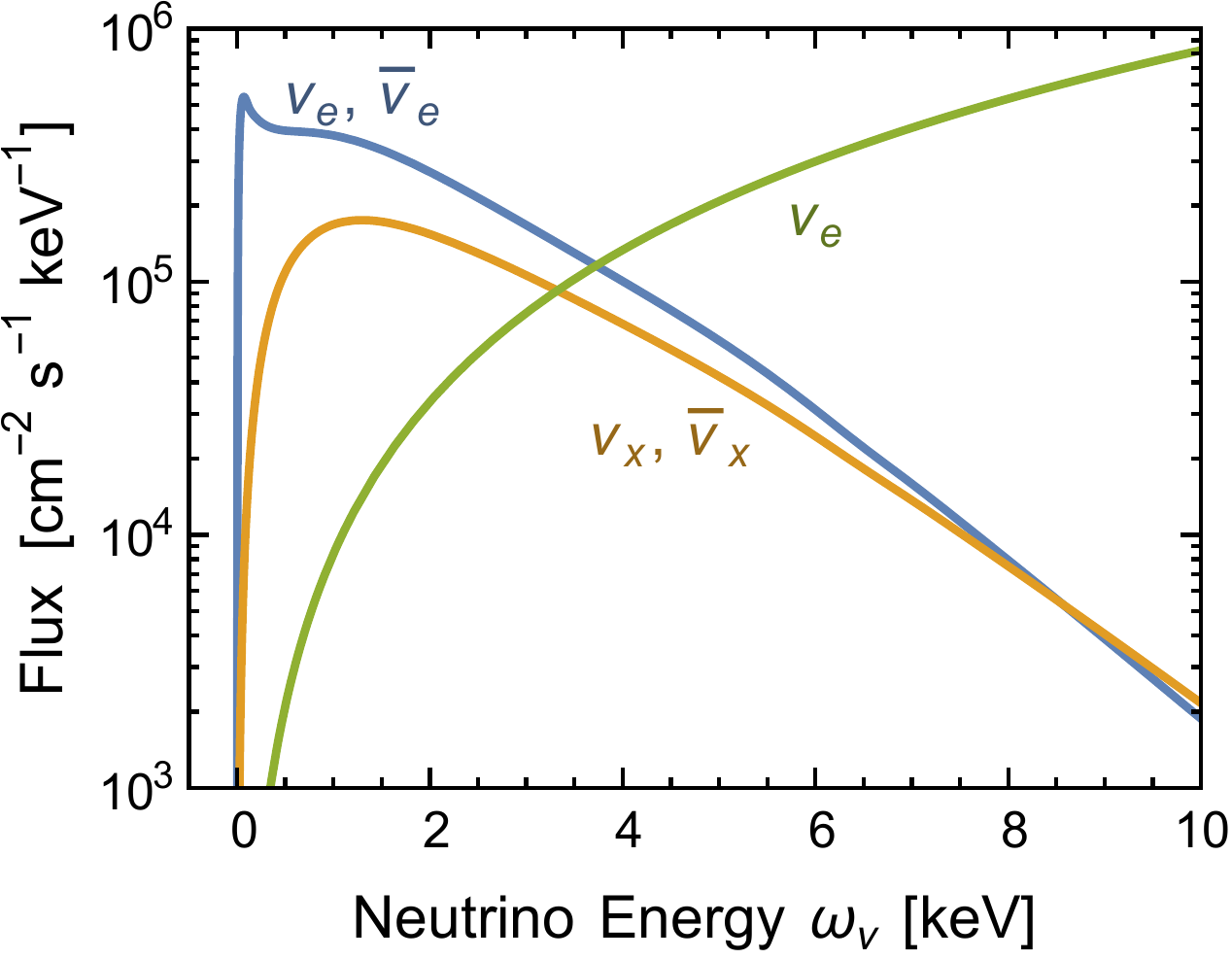}}
\vskip12pt
\hbox to\textwidth{\includegraphics[height=5.8cm]{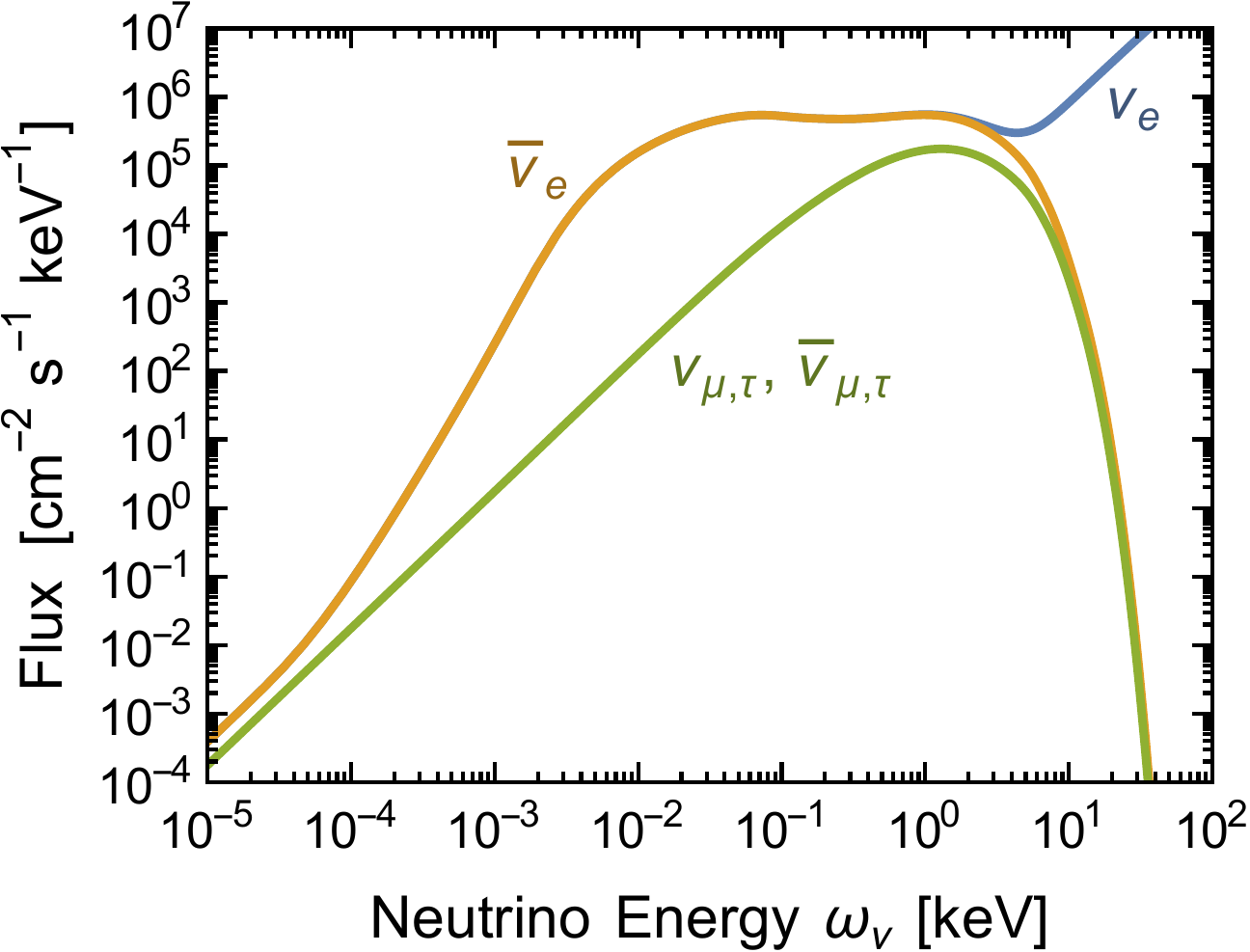}
\hfil\includegraphics[height=5.8cm]{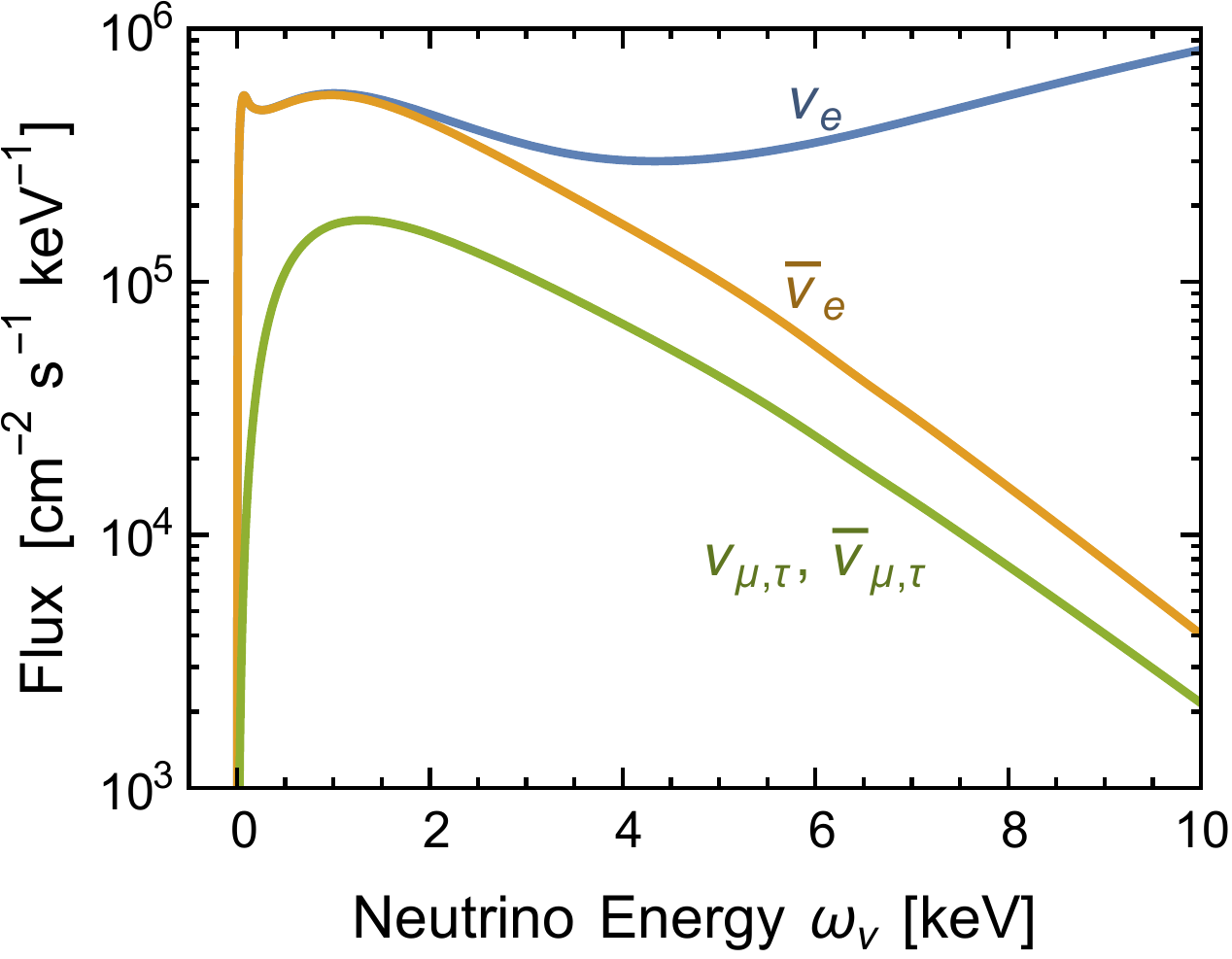}}
\caption{Flavor-dependent solar neutrino fluxes at Earth, ignoring flavor mixing.
{\em Top panels:\/} Thermal flux from vector-current interactions producing only
$\nu_e\bar\nu_e$ pairs, axial-vector emission, producing equal fluxes of every flavor, where
$x$ stands for $e$, $\mu$ or $\tau$, and $\nu_e$ from the nuclear pp reaction.
{\em Bottom panels:\/} $\nu_e$, $\bar\nu_e$, and other single-species fluxes after adding the source
channels.}
\label{fig:total-flavor-flux}
\end{figure}

We may show the same results in a somewhat different form if we observe that the
vector-current processes produce almost exclusively $\nu_e\bar\nu_e$ pairs, whereas
the axial-vector processes produce all flavors in equal measure. In the upper panels of
figure~\ref{fig:total-flavor-flux} we show these total fluxes, where now the coupling constants $\CV^2=0.9263$ and $\CA^2=1/4$ are included. In addition we show the $\nu_e$ flux from the nuclear pp reaction. In the bottom panels
we add the different source channels for every flavor and show the keV-range fluxes for $\nu_e$, $\bar\nu_e$, and each of the other species, still ignoring flavor oscillations.

\subsection{Including flavor mixing}

Flavor-eigenstate neutrinos are mixtures of three different mass eigenstates. After propagating over a long distance, these mass eigenstates effectively decohere, so the neutrino flux arriving at Earth is best described as an incoherent mixture of mass eigenstates. It depends on the nature of a possible detector if these should be re-projected on interaction eigenstates or if these fluxes can be used directly if the detection channel is flavor blind. As we do not know the nature of such a future detector, the neutrino flux in terms of mass eigenstates is the most natural form of presentation.

The axial-current production channel has amplitude $\CA=+1/2$ for $\nu_e$ and $\CA=-1/2$ for the other flavors, yet on the level of the rate (proportional to $\CA^2$) all flavors are produced in equal measure. Therefore, the density matrix in flavor space is proportional to the $3{\times}3$ unit matrix and thus the same in any basis. Without further ado we can think of the axial-current processes as producing mass eigenstates, so in the upper panels of figure~\ref{fig:total-flavor-flux}, the fluxes marked $\nu_x$ and $\bar\nu_x$ can be interpreted as $x$ standing for the mass indices 1, 2 and~3.

The vector-current channels, on the other hand, have the peculiar property of producing almost exclusively $e$-flavored states as discussed earlier, which is also true of the charged-current pp nuclear reaction. These states oscillate in flavor space after production. The relevant oscillation scale is $\omega_{\rm osc}=\Delta m^2/2E=3.8\times10^{-8}~{\rm eV}/E_{\rm keV}$ for the solar mass difference of $\Delta m^2=7.5\times10^{-5}~{\rm eV}^2$ and using $E_{\rm keV}=E/{\rm keV}$. For the atmospheric mass difference $\Delta m^2=2.5\times10^{-3}~{\rm eV}^2$ the oscillation scale is
$\omega_{\rm osc}=1.25\times10^{-3}~{\rm eV}/E_{\rm keV}$. These numbers should be compared with the
matter-induced energy splitting between $\nu_e$ and the other flavors of
$\Delta V=\sqrt{2}\GF n_e= 7.6\times10^{-12}~{\rm eV}$ for the electron density $n_e=6\times10^{25}~{\rm cm}^{-3}$ of the solar center. Therefore, for keV-range neutrinos, the matter effect is very small and neutrino oscillations proceed essentially as in vacuum. The source region in the Sun is much larger than the oscillation length and is far away from Earth, so flavor oscillations effectively decohere long before reaching here. Therefore, we may think of the $e$-flavored channels as producing an incoherent mixture of mass eigenstates. On the probability level, the best-fit mass components of $\nu_e$ are~\cite{Esteban:2016qun}
\begin{equation}
p_1=67.9\%,\qquad p_2=29.9\%,\qquad p_3=2.2\%.
\end{equation}
On this level of precision, these probabilities do not depend on the mass ordering.
The final fluxes in terms of mass eigenstates were shown in figure~\ref{fig:summaryflux}
in the introduction as our main result.

\section{Discussion and summary}
\label{sec:discussion}

We have calculated the solar neutrino flux produced by various thermal processes that produce
$\nu\bar\nu$ pairs with keV-range energies. A proposed dark matter detector for keV-mass sterile neutrinos might find this flux to be a limiting background and conversely, conceivably it could measure this solar flux, although these are somewhat futuristic ideas. Whatever these experimental developments, it is well motivated to provide a benchmark calculation of the thermal solar neutrino flux.

One complication is that there is not a single dominant production channel, but all the processes shown in figure~\ref{fig:processes} are relevant in different ranges of energy. Each of them has its own idiosyncratic issues concerning in-medium many-body effects, yet we think that our flux calculations should be correct on the general 10\% level of precision. Similar uncertainties arise from the variation between different standard solar models.
The only previous study of the keV-range solar neutrino flux in reference~\cite{Haxton:2000xb} stressed the importance of a plasmon enhancement in the photo-production channel, but we think that this result is spurious as argued in section~\ref{sec:pole} and we think that this particular collective effect is not relevant.

It is interesting that the free-bound and bound-bound emission processes on elements heavier than hydrogen and helium produce the dominant flux in the few-keV range. We have calculated this flux taking advantage of the solar opacity calculations available in the literature. The neutrino emissivity is related by a simple phase-space integration to the monochromatic photon emissivity provided by the opacity calculations. Our solar flux based on these processes roughly agrees with that of reference~\cite{Haxton:2000xb} who estimated it by direct calculation on several characteristic elements. Therefore, this flux carries information about the solar metal abundances, quantities of crucial interest in the context of the ``solar opacity problem,'' a point stressed in  reference~\cite{Haxton:2000xb}. As this flux is no longer overshadowed by the spurious plasmon resonance in the photo-production channel, this argument is resurrected by our study, i.e., a measurement of the keV-range flux could provide nontrivial information on the solar metal content.

\section*{Acknowledgments}

We thank Thierry Lasserre for bringing up the question of the
low-energy solar neutrino flux, and Alexander Millar for helpful
discussions.  In Munich, we acknowledge partial support by the
Deutsche Forschungsgemeinschaft through Grant No.\ EXC 153 (Excellence
Cluster ``Universe'') and Grant No.\ SFB 1258 (Collaborative Research
Center ``Neutrinos, Dark Matter, Messengers'') as well as by the
European Union through Grant No.\ H2020-MSCA-ITN-2015/674896
(Innovative Training Network ``Elusives'').  J.R.\ is supported by the
Ramon y Cajal Fellowship 2012-10597 and FPA2015-65745-P
(MINECO/FEDER).

\appendix

\section{Standard solar model}
\label{app:smm}

In the main text we show neutrino fluxes produced from integration over the solar volume. The neutrino production rates are nontrivial functions of several parameters (temperature, density, mass-fraction of each species) which depend on the radial position in the Sun and determine the local plasma properties.
These radial profiles of temperature, density, etc.\ are not directly measured; rather, they are obtained from a solar model. The latter is a theoretical description of the Sun, obtained by evolving certain initial conditions (mass, helium and metal abundances) through a stellar evolution code, which in turn depends on radiative opacities, the treatment of convection, and so forth, to fit the present-day radius, luminosity, and photospheric composition. The latter is the one
present-day boundary condition about which estimations vary the most.

In principle, consistency would require us to use a specific solar model for all processes; this should be a solar model obtained using the opacity code and the abundances exploited for the neutrino bremsstrahlung emission calculation. However, the differences between different solar models are small, introducing uncertainties
in the neutrino emissivity on the order of 10\% perhaps, so we can use more practical criteria for our calculations.
First, we want to use a solar model that not only covers in detail the central core but also the external layers because they rule the low energy flux from T-plasmon decay.\footnote{Note however that the latter is subdominant to the bremstrahlung neutrinos, which are mostly produced in the core.} The Saclay solar model \cite{TurckChieze:2001ye, Couvidat:2002bs} is, in this regard, the most complete one known to us.
This model has the additional interest that it was built to reproduce the sound-speed profile (to a large extent due to the temperature profile) measured by helioseismology by adjusting some parameters (like initial metalicity, opacities) which are not directly measurable.
The model used was built when the surface chemical composition GS98 \cite{Grevesse:1998bj} was suitable to reproduce the helioseismology data with very minor adjustments.

The recent revision of the surface chemical composition AGSS09 \cite{Asplund:2009fu} led to a downward adjustment of abundances, mostly CNO, Ne and refractaries, providing a lower opacity and thus some tension between solar models and helioseismology. This tension has led to a large number of publications but is largely irrelevant for the level of precision that we must assume in our calculations. The difference between GS98 and AGSS09 abundances is 20--40\% in CNO and $\sim 12\%$ in refractaries; both are obtained from spectroscopic observations together with hydrodynamical simulations. We emphasize that the first one agrees with the helioseismic measurements, whereas the second one (which is a more accurate 3D simulation) does not, see reference~\cite{Vinyoles:2016djt} for a recent discussion. This situation constitutes the so-called solar abundance problem.
Differences in sound speed profiles (temperatures) are however only around 1\%.

We prefer to use the GS98 surface composition as it fits better the solar internal structure.
The radial profile of the abundance of several chemical elements is not specified in the Saclay model, so we use instead the abundance profiles
of the GS98 model of reference~\cite{Serenelli:2009yc}.
The profiles of temperature, electron density, plasma frequency, degeneracy parameter $\eta$, Pauli blocking factor, and screening scales for our solar model choice are displayed in the following figures.

On the other hand, by far the most detailed monochromatic opacities publicly available are those of the Opacity Project~\cite{OP}, which are tabulated for different metals and can thus be combined for different solar compositions.
For this reason, they were used in reference~\cite{Redondo:2013wwa} to compute the axion emission in free-bound and bound-bound transitions.
In the same study, three different opacity codes where compared with excellent agreement. Therefore we have used the OP opacities for our calculations in this paper.

\begin{figure}[htbp]
\centering
\hbox to\textwidth{\includegraphics[height=5.6cm]{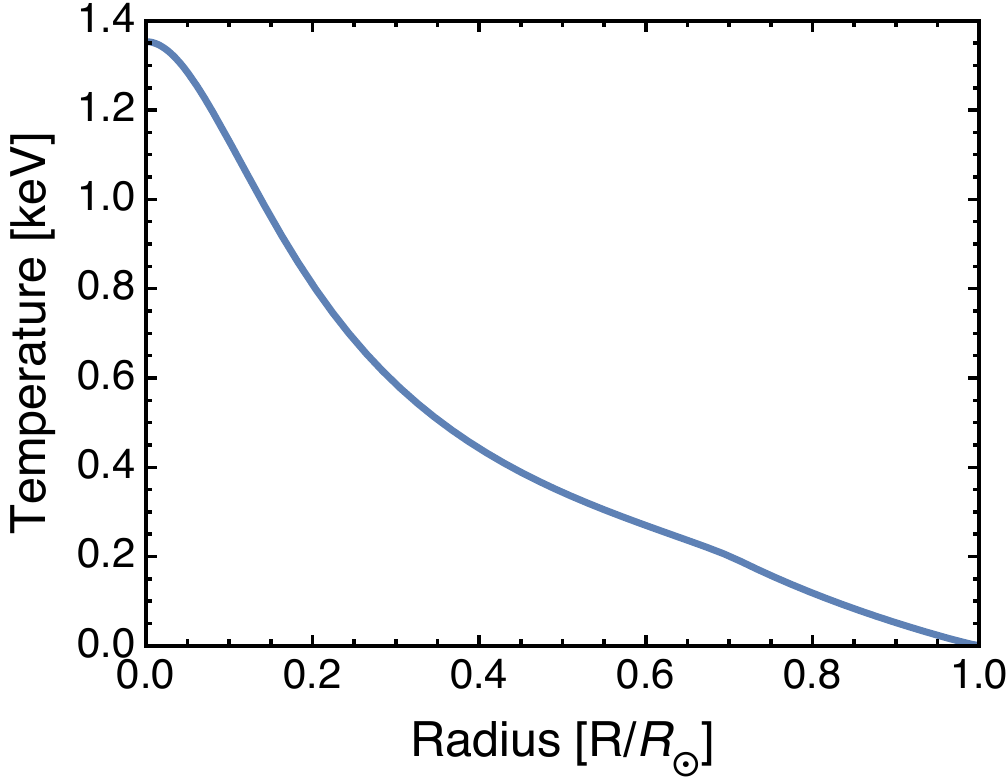}\hfil
\includegraphics[height=5.6cm]{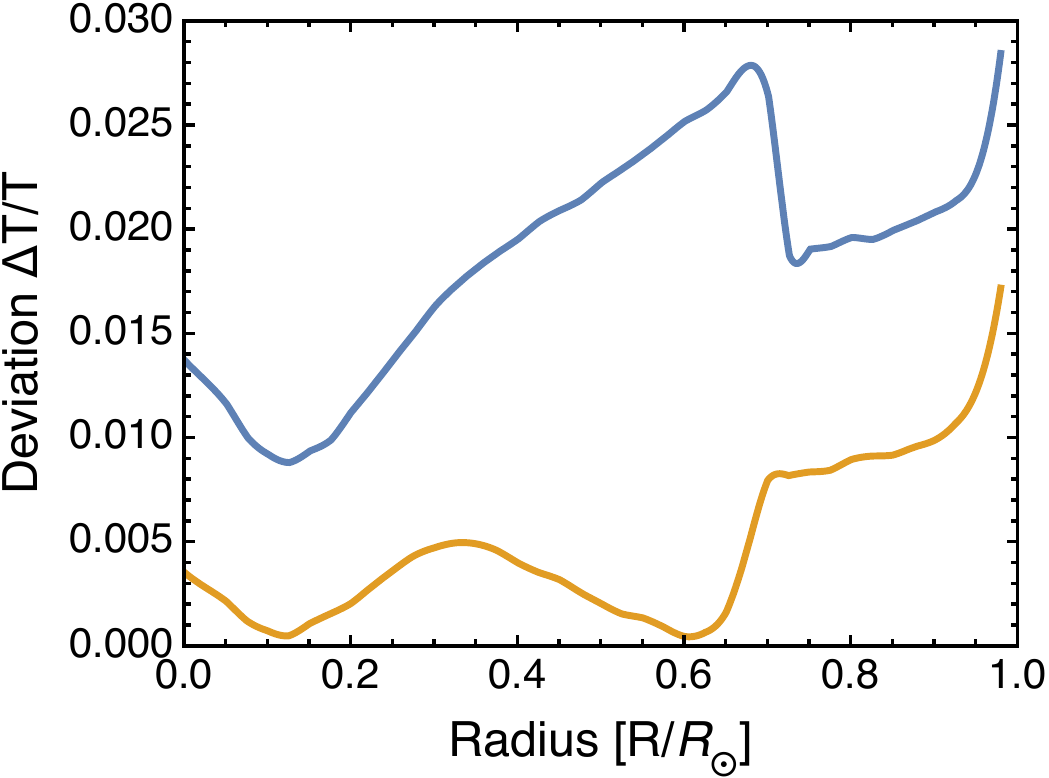}}
\caption{Temperature profile. {\em Left panel:} Saclay model \cite{TurckChieze:2001ye, Couvidat:2002bs}, our standard case. {\em Right panel:}
Relative deviations between solar models.
{\em Blue line:} Saclay model vs.\ model~\cite{Serenelli:2009yc} with AGSS09 abundances. {\em Orange line:} Saclay model vs.\ model~\cite{Serenelli:2009yc} with GS98 abundances.}
\label{fig:tempprof}
\end{figure}

\begin{figure}[htbp]
\centering
\hbox to\textwidth{\includegraphics[height=5.6cm]{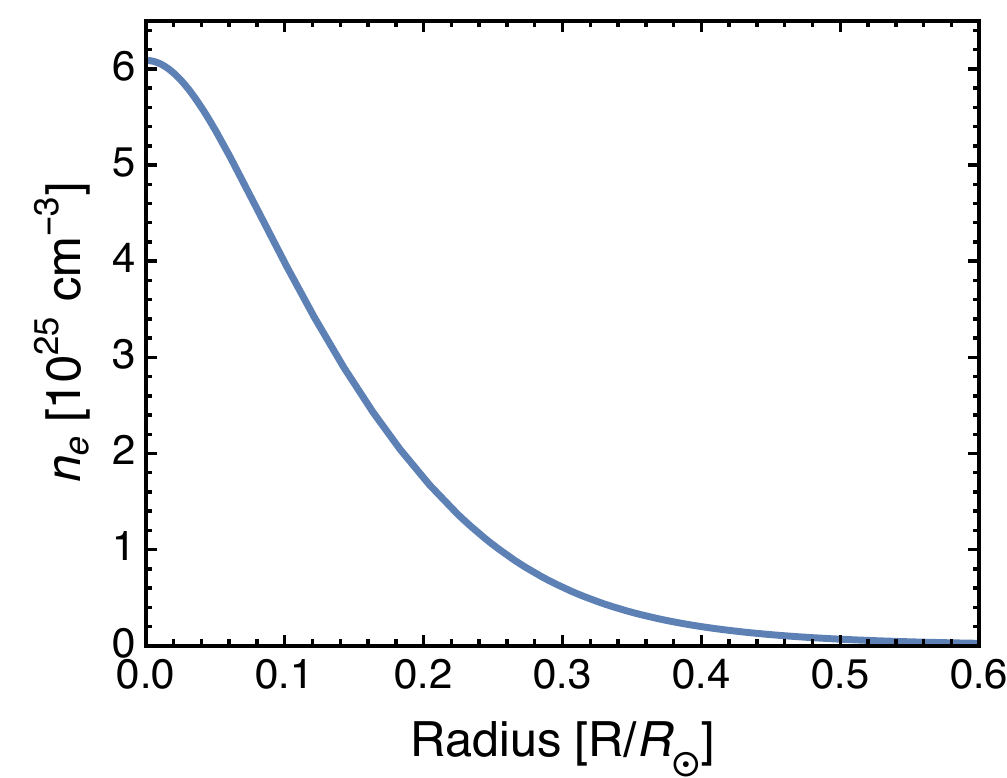}\hfil
\includegraphics[height=5.6cm]{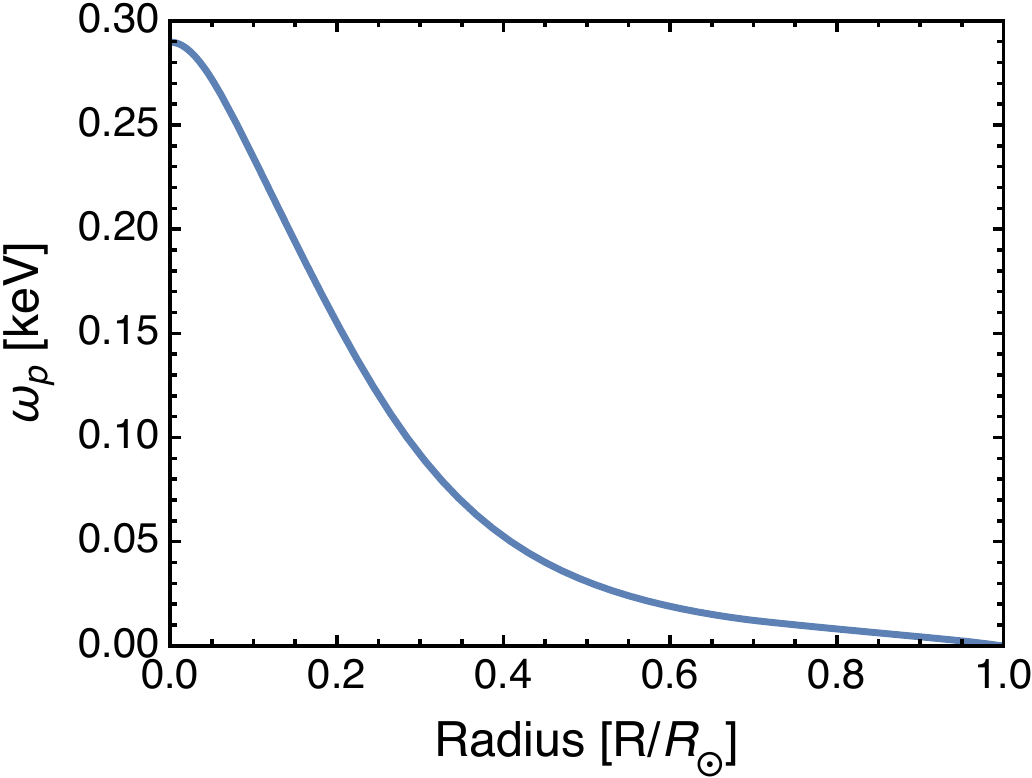}}
\caption{{\em Left panel:} Electron number density. {\em Right panel:} Plasma frequency.}
\label{fig:electrondens}
\end{figure}

\begin{figure}[htbp]
\centering
\hbox to\textwidth{\includegraphics[height=5.6cm]{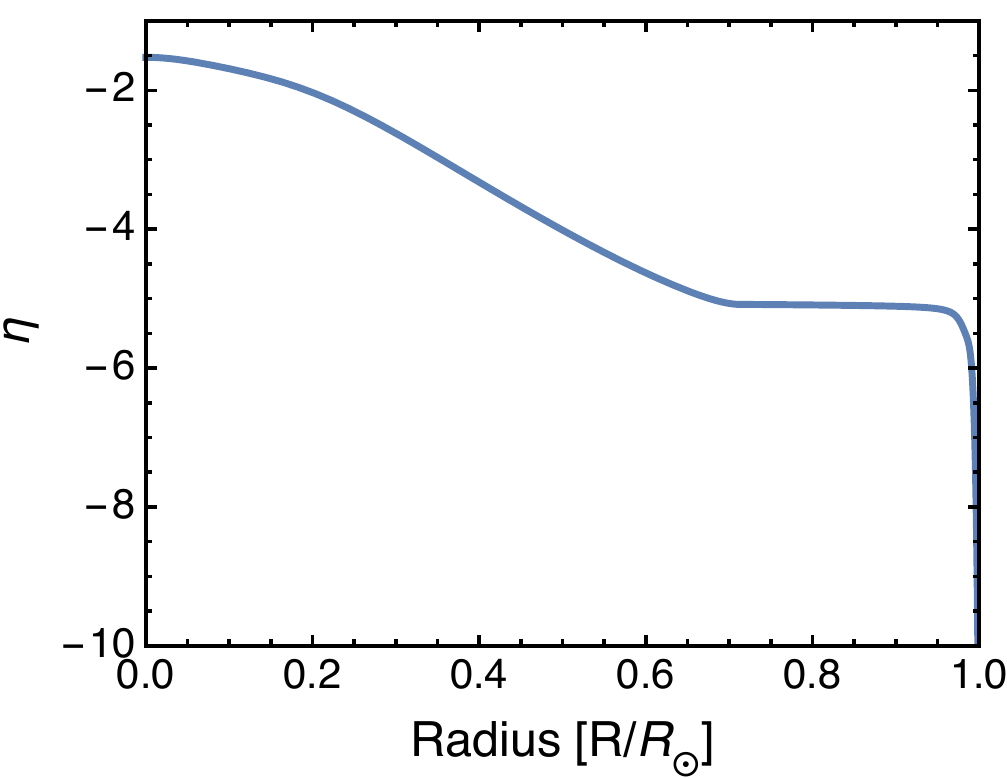}\hfil
\includegraphics[height=5.6cm]{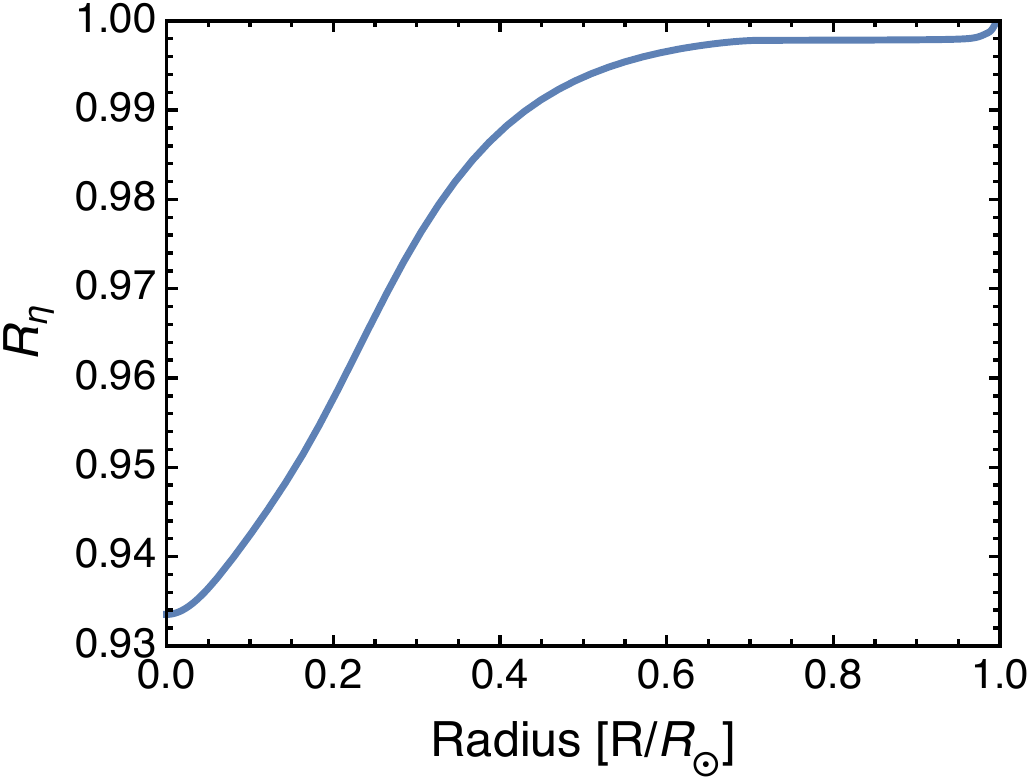}}
\caption{{\em Left panel:} Nonrelativistic degeneracy parameter $\eta$, where $\eta\rightarrow -\infty$ corresponds to a Maxwell-Boltzmann distribution.
{\em Right panel:} Average Pauli blocking factor when electron recoils are small.}
\label{fig:degprof}
\end{figure}

\begin{figure}[htbp]
\centering
\includegraphics[height=5.6cm]{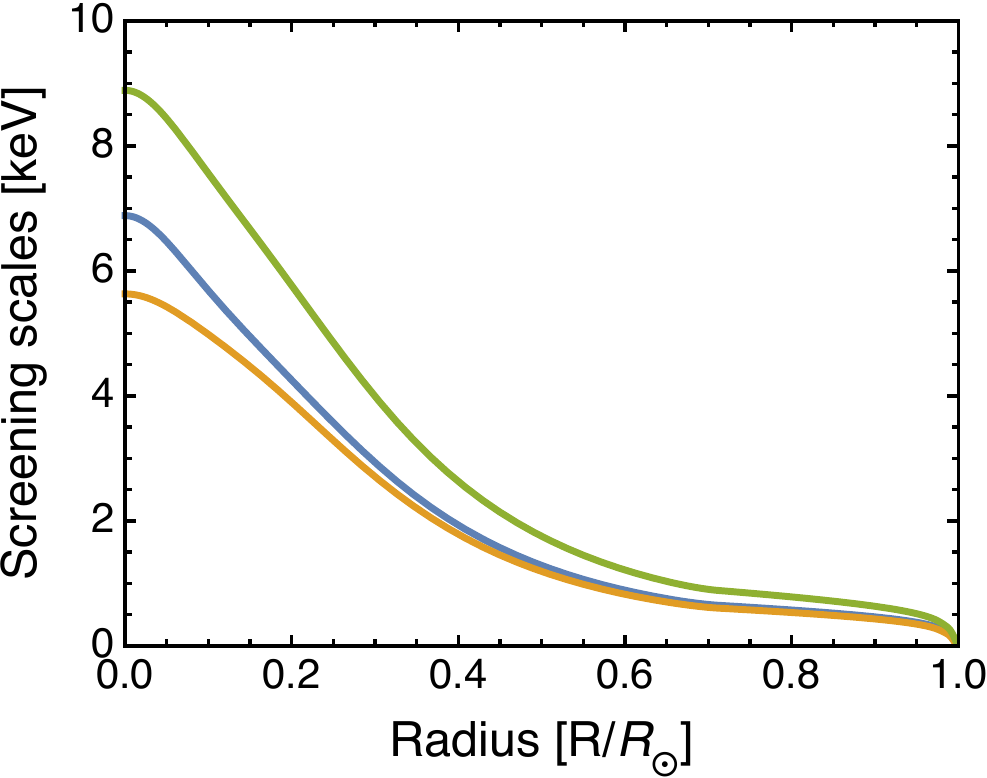}
\caption{Phenomenological screening scales. {\em Orange line:} Electron screening $k_e$. {\em Blue line:} Ion screening $k_i$. {\em Green line:} Total Debye scale
$k_s$.}
\label{fig:screeningscales}
\end{figure}

\begin{figure}[htbp]
\centering
\hbox to\textwidth{\includegraphics[height=6.6cm]{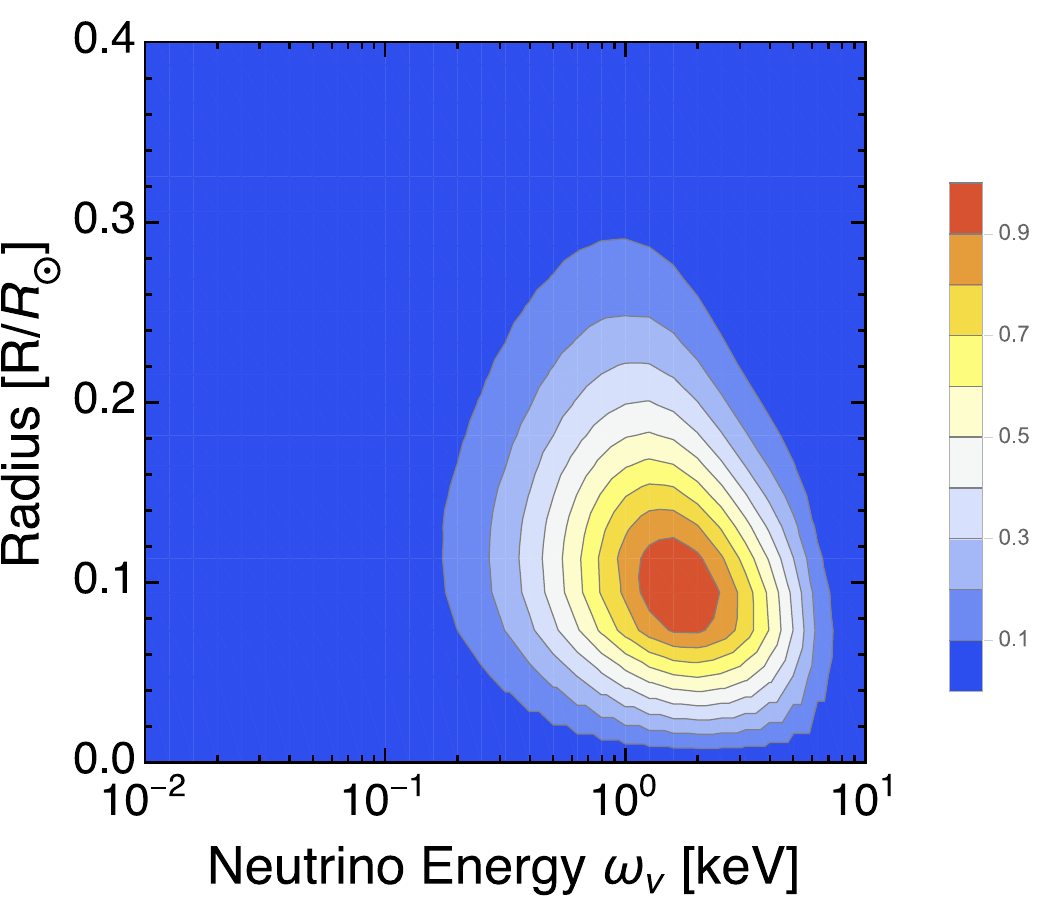}\hfil
\includegraphics[height=6.6cm]{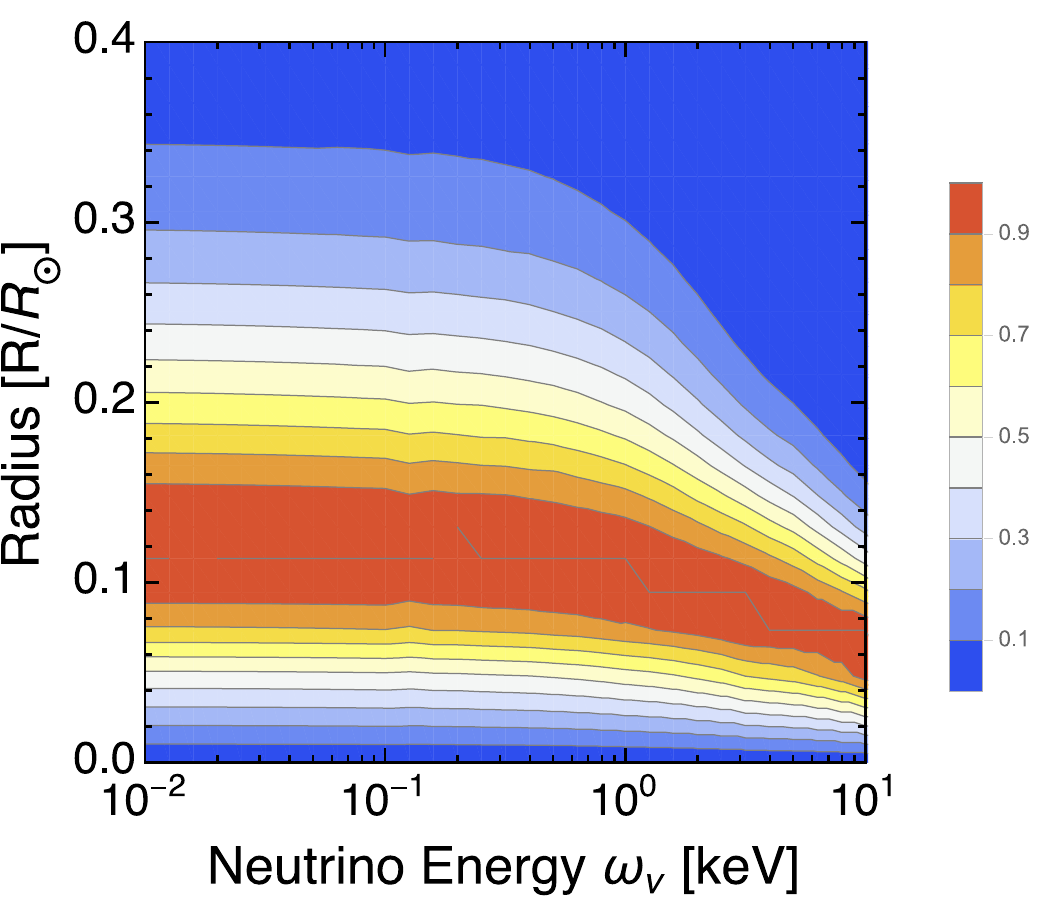}}
\caption{Isocontours of thermal neutrino emission. {\em Left panel:} The total flux (including the geometrical $r^2$ factor) is normalized to 1.
{\em Right panel:} For each neutrino energy, the flux is normalized to 1. The production site of very low energy neutrinos is displaced towards the solar surface, but only slightly.}
\label{fig:radialprofileflux}
\end{figure}

In figures~\ref{fig:tempprof}--\ref{fig:screeningscales} we show the radial variation of various characteristic parameters of our standard solar model. In figure~\ref{fig:radialprofileflux} we show the region of origin of the thermal solar neutrinos calculated in the main text. The left panel shows that most of the flux originates within $0.2~R_\odot$ with a maximum around $0.1~R_\odot$. Notice that emission near the solar center is suppressed by the small volume of this region, i.e., by the geometric $r^2$ factor. In the right panel, we show the same information, but for each
neutrino energy, the flux is normalized to~1. For smaller neutrino energies, the production site is somewhat shifted to larger solar radii.

\pagebreak

\end{document}